\documentclass[
 aps,
 prd,
twocolumn,
superscriptaddress,
frontmatterverbose, 
showpacs,preprintnumbers,
 nofootinbib,
 amsmath,
 amssymb,
floatfix,
]{revtex4-1}
\usepackage[utf8]{inputenc}
\usepackage[english]{babel}
\usepackage{multirow}
\allowdisplaybreaks
\usepackage{amsmath}
\usepackage{amssymb}
\usepackage{booktabs}
\usepackage{graphicx}
\usepackage{placeins}
\usepackage{hyperref}
\usepackage[table]{xcolor}

\graphicspath{{figures/}{../../outputs/}}

\hypersetup{
  colorlinks=true,
  linkcolor=blue,
  citecolor=blue,
  urlcolor=blue
}

\newcommand{\Msun}{M_\odot}
\newcommand{\eps}{\varepsilon}
\newcommand{\deltaratio}{\Delta\varepsilon/\varepsilon_{\rm trans}}
\newcommand{\csq}{c_s^2}
\newcommand{\csssq}{c_{s,{\rm QM}}^2}
 \newcommand{\be}{\begin{equation}}
\newcommand{\ee}{\end{equation}}
\newcommand{\bea}{\begin{eqnarray}}
\newcommand{\eea}{\end{eqnarray}}

\def \W  {\mathcal{W}}
\def \V  {\mathcal{V}}

\begin{document}

\title{Can a slow and strong phase transition in neutron stars relieve major compact-star observation tensions?}

\author{Chen Zhang}
\affiliation{School of Physics Science and Engineering, Tongji University, Shanghai 200092, China}

\date{\today}

\begin{abstract}
Recent anomalous compact-star observations challenge the conventional neutron-star interpretation in complementary ways: HESS J1731--347 and XTE J1814--338 favor unusually small radii at low mass, while the secondary component of GW190814 appears too massive for an ordinary neutron star under GW170817-based maximum-mass inferences.  We examine whether neutron stars with a strong first-order hadron-quark phase transition can address these tensions via the extended stable hybrid branches that arise when the phase conversion is slow compared to radial oscillations, while remaining consistent with GW170817 and NICER constraints.  Using a piecewise-polytropic (PP) benchmark, supplemented by an independent speed-of-sound (CS) parametrization comparison, we find two viable scenarios after a general parameter scan over both the hadronic and hybrid branches.  Scenario 1 realizes an all-at-once solution: the same equation of state has a pure hadronic branch compatible with both the GW190814-scale mass and HESS J1731--347, while its slow stable hybrid branch reaches XTE J1814--338.  Scenario 2 retains the GW190814-scale hadronic branch but reaches XTE J1814--338 and HESS J1731--347 on slow stable hybrid branches in very different transition-strength regimes.

\end{abstract}

\maketitle

\section{Introduction}
\label{sec:intro}

Neutron stars (NSs) are natural laboratories for strongly interacting matter compressed beyond
nuclear saturation density, a regime inaccessible to terrestrial experiments.  Modern astrophysical observations of NSs thus provide a way to probe the properties of dense nuclear matter and constrain the equation of state (EOS).  Gravitational-wave (GW) events observed by LIGO/Virgo, particularly GW170817, place upper bounds on tidal
deformability in binary inspirals that exclude stiff EOSs at intermediate densities~\cite{Abbott2017GW170817,Hinderer2008,Hinderer2010,Postnikov2010};
recent NICER pulse-profile modeling of x-ray observations provides direct mass-radius information
\cite{Miller2019NICERJ0030,Riley2019NICERJ0030,Miller2021NICERJ0740,Riley2021NICERJ0740,
Choudhury2024NICERJ0437,Rutherford2024NICEREOS} that maps directly onto constraints on the EOS.  Growing evidence indicates a phase transition (PT) from hadronic matter (HM) to quark matter (QM) may exist inside massive neutron stars~\cite{Annala:2019puf,Annala:2023cwx,Fujimoto:2022ohj,Komoltsev:2024lcr,PhysRevD.107.023012}, forming so-called hybrid stars.

However, some recent compact-star observations seem anomalous in the conventional NS picture:
GW190814 contained a $2.59^{+0.08}_{-0.09}\,\Msun$ secondary \cite{Abbott2020GW190814} which is in tension with maximum-mass and
tidal constraints inferred from GW170817  if it was a nonrotating neutron star~\cite{Margalit2017,Rezzolla2018,Ruiz2018,Shibata2019}.  In contrast,
HESS J1731--347 and XTE J1814--338 have been inferred to occupy unusually low-mass, small-radius regions of the
$M$-$R$ plane,
with $M=0.77^{+0.20}_{-0.17}\, M_\odot$, $R=10.4^{+0.86}_{-0.78}$ km for HESS J1731--347
\cite{Doroshenko2022} and $M=1.21^{+0.05}_{-0.05}\, M_\odot$, $R=7.0^{+0.4}_{-0.4}\, \rm km$ for XTE J1814--338
\cite{Kini2024}.  These inferences have motivated hybrid-star and other compact-object
interpretations
\cite{LaskosPatkos2024HESS,LaskosPatkos2025XTE,Yang:2024ycl,Mariani2024HESS,Mariani:2025nud,Zhou:2026exw,Dong:2026kpb}.  However,  they are difficult to reconcile with large mass measurements in a
single EOS sequence: the pressure needed for a
GW190814-scale maximum mass tends to enlarge ordinary-mass radii, while an EOS soft enough for HESS J1731--347 and XTE J1814--338
usually struggles to regain the required high-density support.

The mass-radius inferences for HESS J1731--347 and XTE J1814--338
should be interpreted with caution.  For HESS J1731--347, the uniform carbon
atmosphere model adopted in Ref.~\cite{Doroshenko2022} is supported by the
absence of pulsations and the observed blackbodylike spectra, with
alternative atmosphere models explored but found less consistent with the
data.  Related analyses of central compact objects have shown that a
uniform-temperature carbon atmosphere is not a unique spectral description,
so the inferred contour remains model dependent~\cite{AlfordHalpern2023}.
For XTE J1814--338, the single hot spot model of
Ref.~\cite{Kini2024} successfully reproduces the fundamental
fractional amplitudes and incorporates improvements over earlier analyses,
though some discrepancies remain at higher harmonics, motivating future
refinements with multihot spot models.
For GW190814, the secondary mass lies in the lower mass gap between
the heaviest known neutron stars and the lightest astrophysical black holes.
The absence of a detectable tidal signature or electromagnetic counterpart
leaves both interpretations open.  A neutron-star reading would make it the
heaviest NS yet and require $M_{\rm TOV}$ at the quoted scale, which is the
tension this work addresses~\cite{Abbott2020GW190814,Tews2021GW190814,LVK2026GWTC5Population}.
We therefore use the quoted contours and mass as deliberately stringent
benchmarks and return in Sec.~\ref{sec:summary} to how relaxing each benchmark
changes the two scenarios.

Strong first-order phase transitions can decouple these requirements
\cite{Dexheimer2018PhaseTransitions,Alford2005HybridMasquerade,Alford2013CSS,Seidov1971,Schertler2000}.
The decisive point is dynamical.  For rapid
conversions (compared to the radial oscillation timescale) at the phase boundary, the usual turning-point stability criterion ($\partial M/\partial P_c>0$) remains a good guide.
However, for slow conversions, configurations can remain stable past the local or global maximum-mass point 
\cite{Haensel1989,Pereira2018,Mariani2019,Lugones2021PhaseConversions,Lugones2023SlowStable,
Goncalves2022TwinModes,Rau2023IntermediateConversion,Mariani:2026xdh}, with a general feature that a stronger phase transition (i.e., larger density discontinuities) generates a more extended stable branch of the so-called slow stable hybrid stars (SSHS) that can be formed from early phase transitions triggered by quantum or thermal nucleations in realistic astrophysical scenarios such as protoneutron stars or hot postmerger remnants~\cite{Lugones:2021bkm}.  This extended slow stable branch can therefore explore the small radii regime unavailable to the pure hadronic sequence. Besides, this slow stable branch picture can make these quoted maximum-mass limits set by various post-GW170817 analyses not directly applicable as priors, since these analyses did not consider this branch and potential violations of their assumptions by the associated very strong phase transitions~\cite{han2019tidal,hoyos2022holographic,Zhang2026SelfBound}. 

The physics determining whether the transition is slow or rapid is subject to substantial uncertainty due to the nonperturbative nature of strong dynamics. Although the interface surface tension—which hinders the transition—is notoriously model-dependent, we can still anticipate a slow phase transition if the flavor compositions of the two phases differ, since this would necessitate a high-order weak interaction process.

Reference~\cite{LaskosPatkos2025XTE} showed that two fixed hadronic EOSs (APR and DD2) matched to a constant-sound-speed (CSS) parametrization of the quark matter phase can reach the XTE J1814--338 region while meeting the lower bound on the maximum gravitational mass, $M_{\rm TOV}\gtrsim 2M_\odot$~\cite{Demorest2010,Antoniadis2013,Cromartie2020}, once slow conversion is allowed.  Reference~\cite{Mariani:2025nud} extended this analysis but considered only a limited region of parameter space and did not take GW190814 into account.  In contrast, we ask the important question of whether more generally parametrized, crust-bearing EOSs can simultaneously
retain a hadronic branch capable of supporting a GW190814-scale mass together with XTE J1814--338 and/or HESS J1731--347.  To the authors' knowledge, no existing neutron-star study has addressed GW190814 together with XTE J1814--338 or HESS J1731--347 within a single EOS framework, let alone all three observations simultaneously.  We also keep the standard multimessenger context in view, including GW170817 and
NICER mass-radius constraints for PSR J0030+0451, PSR J0740+6620, and PSR J0437--4715
\cite{Miller2019NICERJ0030,Riley2019NICERJ0030,Miller2021NICERJ0740,Riley2021NICERJ0740,
Choudhury2024NICERJ0437,Rutherford2024NICEREOS}. 

Note that this study also differs from our previous study of self-bound
hybrid-star scenarios \cite{Zhang2026SelfBound} that were based on unconventional phase
transitions~\cite{HybQS,Zhang:2022pse,Zhang:2023zth,Negreiros:2024cvr,Zhang:2023szb,Teruya:2025tpx}
inside self-bound compact stars~\cite{Haensel:1986qb,Alcock:1986hz, Zhang:2019mqb,Ren:2020tll,Cao:2020zxi,Yuan:2022dxb,Zhang:2020jmb,Zhang:2021iah,Pretel:2024pem,Zhou:2024syq,
Xie:2025sth,Miao:2024qik,Lai:2009cn,Zhang:2023mzb, Li:2026skf,Yuan:2024hge,Yuan:2026yae} composed of quark matter~\cite{Bodmer:1971we,Witten:1984rs, Farhi:1984qu,Holdom:2017gdc} or quark-cluster
matter~\cite{Xu:2003xe}, where the small-radius behavior at low mass that helps accommodate HESS J1731--347 is aided by the finite-density self-bound star surface.   

This paper is organized as follows.  Section~II lays out the complete
EOS assembly and scan construction, including the parameter grids.
Section~III presents the results in hierarchical blocks: the figure map and
the two observational scenarios (including tidal response), the effects of
transition mass and jump strength, the posttransition stiffness comparison,
central densities and the pQCD domain, and the PP-CS comparison.
The Appendixes are reserved for the stellar-structure and radial-stability
equations, the accelerated end point-search method, and the hadronic
sound-speed diagnostics.

\section{Model and Method}
\label{sec:hadronic}

\begin{table*}[!t]
\centering
\caption{Construction of the EOS ensembles used in the figures.  Braces
list the discrete values of the archived grids; intervals give the parent
domains.  All hadronic models are tested against Eq.~(\ref{eq:fermi-proxy}) and
$M_{\rm TOV}>2.55\,\Msun$ after causal treatment.}
\label{tab:scan-design}
\arrayrulecolor{black}
\begin{tabular}{|l|l|}
\hline
\parbox[t]{3.2cm}{Stage/model} &
\parbox[t]{12.8cm}{Parameters and sampled values} \\ \hline
\parbox[t]{3.2cm}{Low density} &
\parbox[t]{12.8cm}{BPS for $n\leq0.5n_0$; soft or stiff $\chi$EFT edge up to
$n_{\rm cEFT}=1.1n_0$} \\ \hline
\parbox[t]{3.2cm}{PP core, four- and eight-point grids} &
\parbox[t]{12.8cm}{Four or eight equally spaced values over the same parent ranges:
$1\leq\Gamma_1\leq4.5$, $0<\Gamma_2\leq8$,
$0.5\leq\Gamma_3\leq8$,
$1.5\leq\rho_{12}/\rho_0\leq8$, and
$\rho_{12}\leq\rho_{23}\leq8.5\rho_0$.  The $\Gamma_2=0$ end point of the
underlying grids is excluded before selection.} \\ \hline
\parbox[t]{3.2cm}{CS core} &
\parbox[t]{12.8cm}{$a_1=\{1.2,1.5,1.8,2.0\}$,
$a_2=\{4,6,8,10,12\}$,
$a_3=\{2,4,6,8,12\}$ with $a_3\leq2a_2$,
$a_4=\{3,6,9,12\}$,
$a_5=\{0.4,0.7,1.0,1.4\}$,
$a_7=\{0.7,0.85,1.0\}$} \\ \hline
\parbox[t]{3.2cm}{CSS transition} &
\parbox[t]{12.8cm}{$M_t=\{2.51,2.60\}\,\Msun$;
$1\leq\Delta\eps/\eps_{\rm trans}\leq50$;
$c_{s,\rm QM}^2=1$ (main scan) or $0.8$ (sensitivity scan)} \\ \hline
\end{tabular}
\end{table*}

\subsection{EOS assembly}
\label{subsec:eos-assembly}

We construct conventional neutron stars from a BPS crust ($n\leq0.5\,n_0$) joined to
either the lower (softer) or upper (stiffer) edge of the microscopic $\chi$EFT band,
which extends up to $n_{\rm cEFT}=1.1\,n_0$
\cite{Baym1971,Hebeler2013,Greif2019}, where $n_0$ is the saturation density.
Specifically, the outer-crust EOS is the BPS EOS of Ref.~\cite{Baym1971},
and the soft and stiff microscopic bands are, respectively, the lower and upper
edges of the tabulated chiral-EFT uncertainty band constructed in
Ref.~\cite{Hebeler2013} and used in the sound-speed implementation of
Ref.~\cite{Greif2019}.  This BPS+$\chi$EFT construction provides the common low-density baseline for all hadronic core extensions used below.  We first use
the widely adopted piecewise-polytropic (PP) model~\cite{Hebeler2013,Read2009}.  In
each segment
\be
P=K_i\rho^{\Gamma_i},
\ee
where the mass density $\rho=m_n n$, energy density $\eps(\rho)=(1+a_i)\rho+{K_i} \rho^{\Gamma_i}/(\Gamma_i-1)$ with $a_{i} =  \epsilon(\rho_{i-1})/\rho_{i-1} - 1  - K_{i} \rho_{i-1}^{\Gamma_{i} - 1}/(\Gamma_{i} - 1) $ and $K_{i+1} = P(\rho_i)/\rho_i^{\Gamma_{i+1}}$
as determined from the usual continuity conditions at the segment
interfaces, while $a_0=0$ from the requirement $\displaystyle \lim_{\rho \rightarrow 0} \epsilon/\rho = 1$ and $K_1$ is determined by the continuity at  the crust-core interface at $n_{\rm cEFT}$.  At the first point where
$c_s^2\to1$, corresponding to $\eps_{\rm limit}$, we apply the causal extension of Hebeler et al.~\cite{Hebeler2013}
\begin{equation}
  P(\eps) = P_{\rm limit} + (\eps-\eps_{\rm limit}),
  \qquad \eps>\eps_{\rm limit},
  \label{eq:causal-ext}
\end{equation}
which ensures that the energy density, pressure, and speed of sound are continuous at $\eps_{\rm limit}$, with $c_s^2=1$ for $\eps> \eps_{\rm limit}$.

The $c_s^2=1$ continuation in Eq.~\eqref{eq:causal-ext} is a causal
completion: it replaces only the acausal tail of a sampled polytrope, leaves
the lower-density EOS unchanged, and provides the maximally stiff upper
envelope for the PP scan~\cite{RhoadesRuffini1974,Hebeler2013}, not a
microscopic hadronic prediction.  Models whose GW190814-scale support requires
a long causal plateau are therefore explicitly extremal.

As an independent hadronic core model for a robustness check, we also use
the speed of sound (CS) parametrization~\cite{Greif2019,LaskosPatkos2025SoundBounds,Zhou:2026exw},
\begin{equation}\label{eq:cs2}
\begin{split}
  \csq(x)
  &= a_1 \exp\left[-\frac{1}{2}\left(\frac{x-a_2}{a_3}\right)^2\right]
  + a_6 \\
  &+ \frac{a_7-a_6}{1+\exp[-a_5(x-a_4)]},
  \end{split}
\end{equation}
where $x=\eps/\eps_0$ and $\eps_0=m_n n_0$.  The six scan parameters are
$(a_1,a_2,a_3,a_4,a_5,a_7)$; the offset $a_6$ is fixed by the crust-core matching condition.~\footnote{Note that we do not impose the condition $a_6<0$ as in Ref.~\cite{Zhou:2026exw}, since we do not use the CS model itself for phase transition. }
The pressure is obtained by integrating $dP/d\eps=\csq$ from the interface.  Following the
causal implementation used in recent CS-parametrization studies, whenever the raw formula gives
values outside the physical interval we tabulate
$\csq(x)\rightarrow\min[\max(c_{s,\mathrm{raw}}^{2}(x),0),1]$, so each pressure table has
$0\leq c_s^2\leq1$ by construction~\cite{Zhou:2026exw}.

Above the transition pressure $P_{\rm trans}$, we attach a commonly used CSS parametrization of the quark-matter phase through a Maxwell
construction:
\begin{equation}
  \eps(P)=\eps_{\rm trans}+\Delta\eps+
  \frac{P-P_{\rm trans}}{\csssq},
  \qquad P>P_{\rm trans},
  \label{eq:css}
\end{equation}
where $\eps_{\rm trans}$ is the hadronic energy density at the transition~\cite{Alford2013CSS}, and $\csssq$ is the squared speed of sound in the quark-matter core.  

The main scan uses $\csssq=1$, the causal-limit CSS member that
maximizes posttransition support~\cite{Alford2013CSS,RhoadesRuffini1974}; it
is an extremal existence test for connecting a GW190814-scale transition to
the small-radius XTE window, not a perturbative-QCD prediction.  Lower
$\csssq$ shortens the slow stable branch, as in a complete $c_s^2=0.1$--$1$
scan of self-bound hybrid stars~\cite{Zhang2026SelfBound} used here as a
qualitative match; the $\csssq=0.8$ repeat probes sensitivity near that
limit only.

Stellar sequences are obtained from the Tolman-Oppenheimer-Volkoff equations given in
Appendix~\ref{app:stability-equations}.  The transition is placed at selected masses
$M_t$ on the hadronic branch with varying $\deltaratio$ and $\csssq$.  

We devise a residual method (RM) for locating the end points of slow stable branches in Appendix~\ref{app:method1}, saving us from the time-consuming conventional shooting method in the oscillation frequency domain, by directly putting $\omega^2=0$ in the radial oscillation equations [Eqs.~\eqref{eqXi} and~\eqref{eqDP}] and only look for the center pressure point where the surface residual [Eq.~\eqref{eq:method1-residual}] vanishes. This accelerates end point finding on an order-of-magnitude timescale.
We verified that the resulting end points agree with those obtained from the conventional shooting calculation.

\begin{figure*}[t]
  \centering
  \begin{minipage}[t]{0.49\textwidth}
    \centering
    \includegraphics[width=\linewidth]{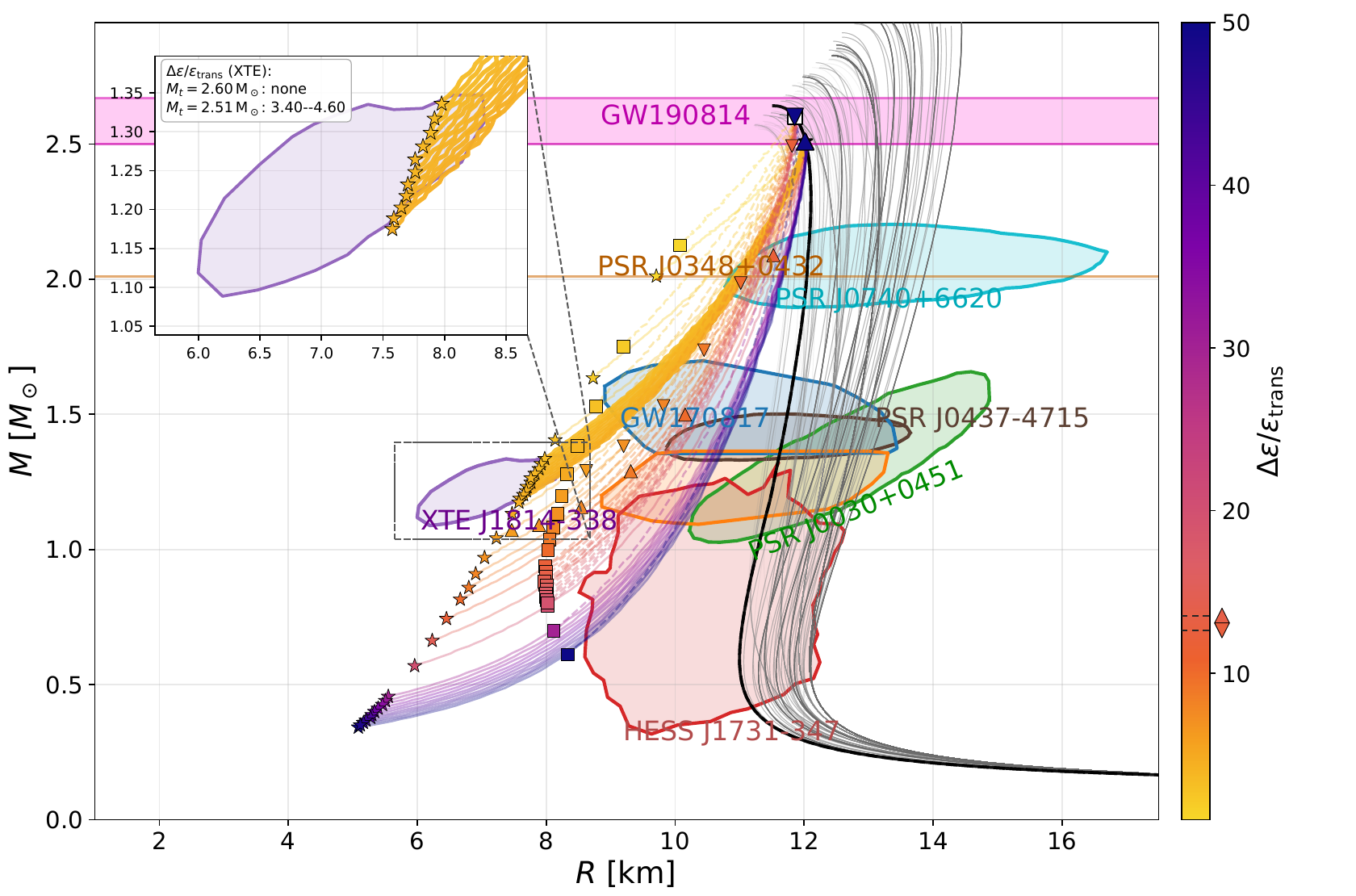}\\[-0.4em]
    \textbf{(a)}
  \end{minipage}\hfill
  \begin{minipage}[t]{0.49\textwidth}
    \centering
    \includegraphics[width=\linewidth]{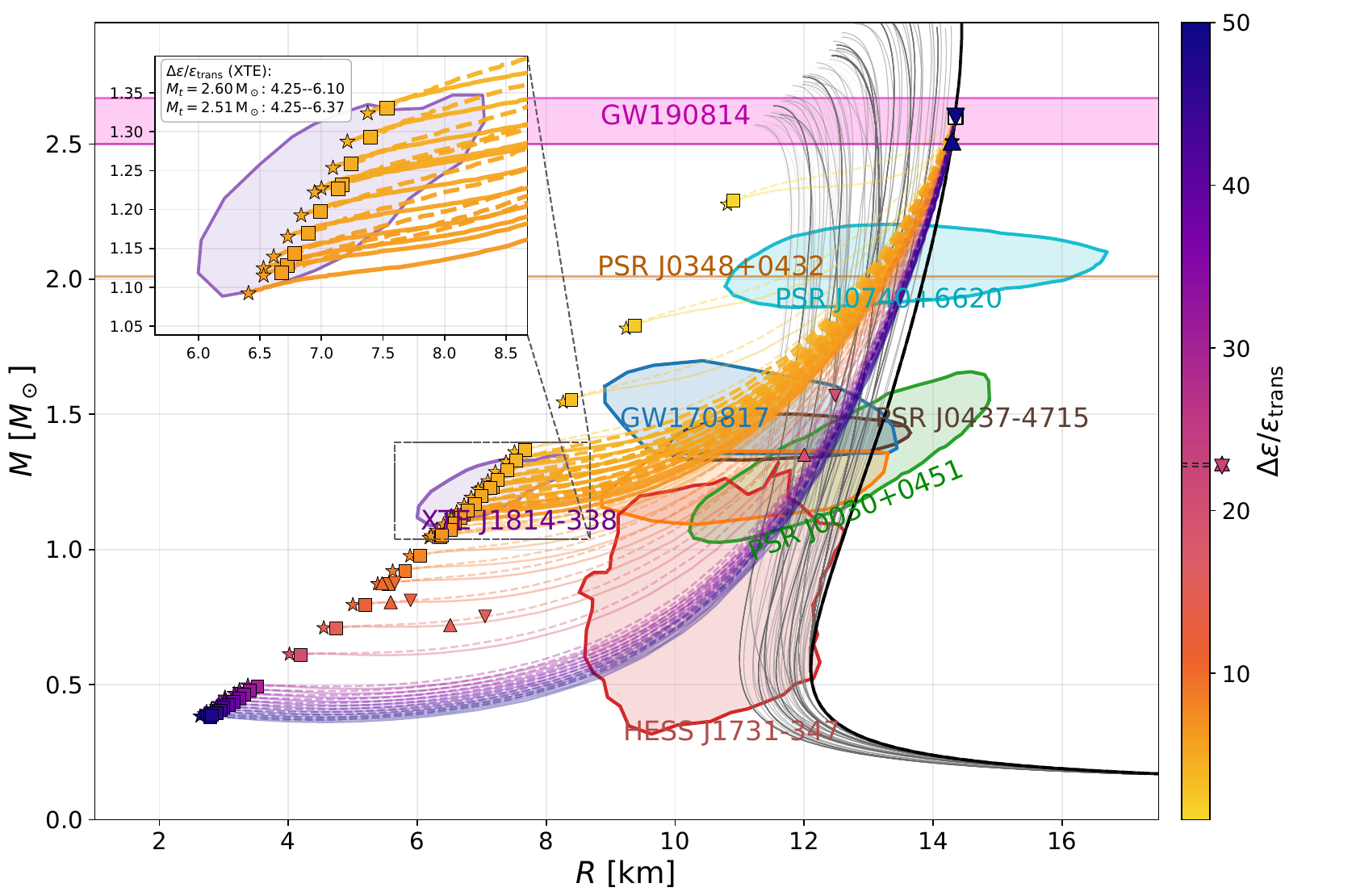}\\[-0.4em]
    \textbf{(b)}
  \end{minipage}
  \caption{$M$--$R$ scans for the BPS+$\chi$EFT(soft)+PP benchmark after the
FLT and $M_{\rm TOV}>2.55\,\Msun$ filters.  Colored curves show the slow stable
hybrid branches with $\csssq=1$ built from the selected pure-hadronic
sequences (black), while thin gray curves show the other passing hadronic
scan.  Panels (a) and (b) use representative
hadronic branches of moderate-$M_{\rm TOV}$ ($2.64\,\Msun$) and highest-$M_{\rm TOV}$ ($3.22\,\Msun$), with corresponding EOS parameters
$(\Gamma_1,\Gamma_2,\Gamma_3;\rho_{12}/\rho_0,\rho_{23}/\rho_0)
=(1,5.333,0.5;1.5,8.5)$ and $(4.5,8,0.5;1.5,6.167)$, respectively.
Line style and end point symbols distinguish the transition mass: solid curves
with stars correspond to $M_t=2.51\,\Msun$, while dashed curves with boxes
correspond to $M_t=2.60\,\Msun$.  The inset gives the XTE-intersecting
$\Delta\eps/\eps_{\rm trans}$ range.
Filled triangles mark $n_c=n_{\rm pQCD}=40n_0$ on the slow stable
branch (upright for $M_t=2.51\,\Msun$, inverted for $M_t=2.60\,\Msun$),
stacked on the transition symbol when that density is already reached at
the Maxwell jump; dashed color-bar ticks give the critical
$\Delta\eps/\eps_{\rm trans}$ of Eq.~\eqref{eq:npqcd-crit}
(Sec.~\ref{subsec:pqcd}).}
  \label{fig:hebeler-xte}
\end{figure*}

\subsection{Scan construction, parameter ranges, and selection}
\label{subsec:scan-selection}

The scan is assembled in three ordered stages: select a low-density
BPS+$\chi$EFT matching, generate and filter a hadronic core, and then attach
the CSS phase at prescribed points of each retained hadronic sequence.  Table~I
collects the complete parameter domains and grids so that every ensemble used
in Sec.~III can be reconstructed from this subsection.

We require all hadronic sequences to satisfy two baseline conditions before the phase
transition is added.  First, we apply a common low-density Fermi-liquid-theory (FLT) constraint, often used in
the CS-parametrization literature~\cite{Schwenk2003,SchwenkFriman2004,Greif2019,Zhou:2026exw}.
For the polytropic tables this condition is evaluated after converting the tabulated EOS to
$c_s^2=dP/d\eps$.  The FLT constraint is
\begin{equation}
 \csq(1.5\,\eps_0)<0.163.
  \label{eq:fermi-proxy}
\end{equation}
Second, we require $M_{\rm TOV}>2.55\,\Msun$ to accommodate the secondary object of GW190814.

\begin{figure*}[t]
  \centering
  \begin{minipage}[t]{0.49\textwidth}
    \centering
    \includegraphics[width=\linewidth]{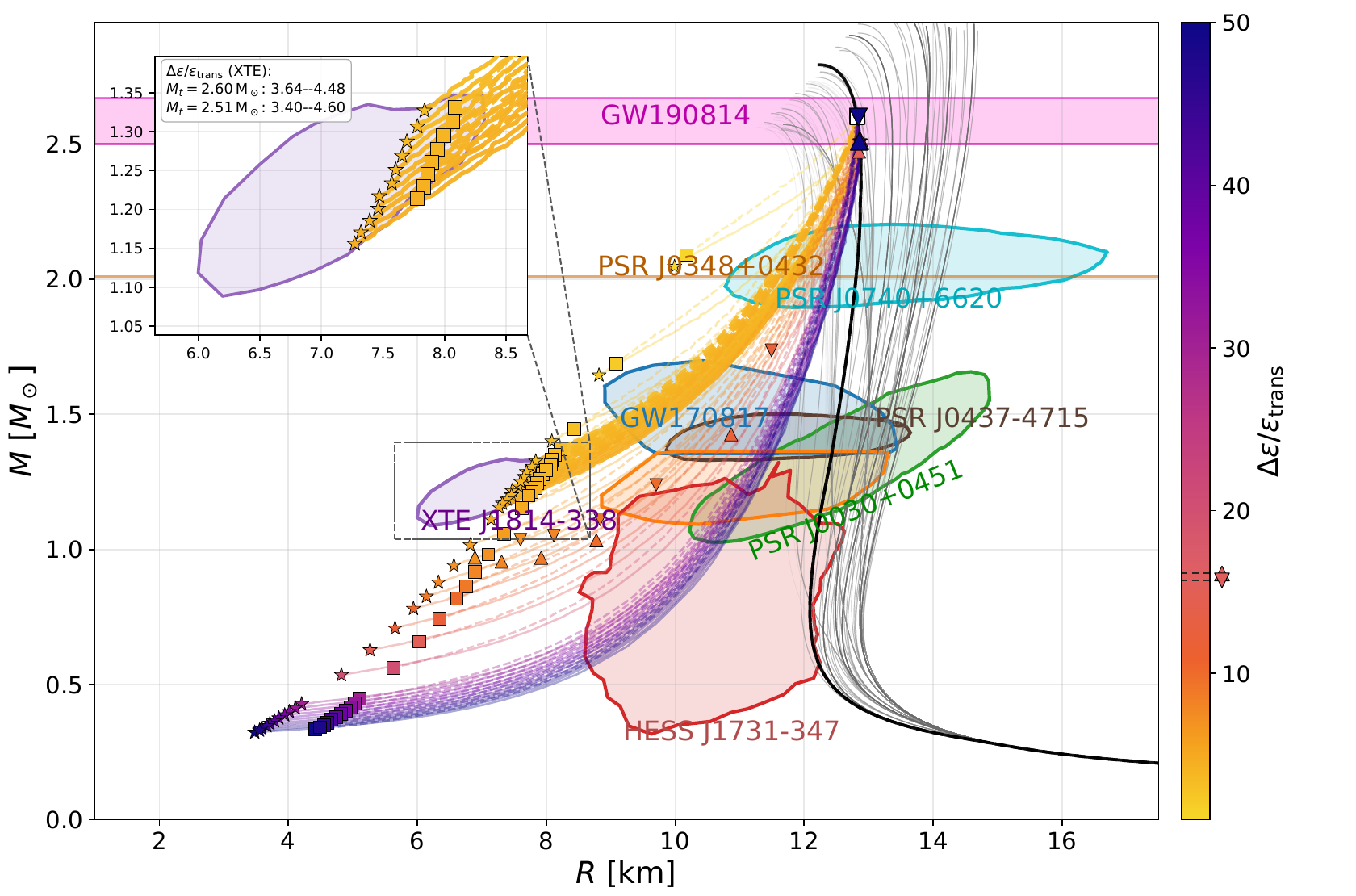}\\[-0.4em]
    \textbf{(a)}
  \end{minipage}\hfill
  \begin{minipage}[t]{0.49\textwidth}
    \centering
    \includegraphics[width=\linewidth]{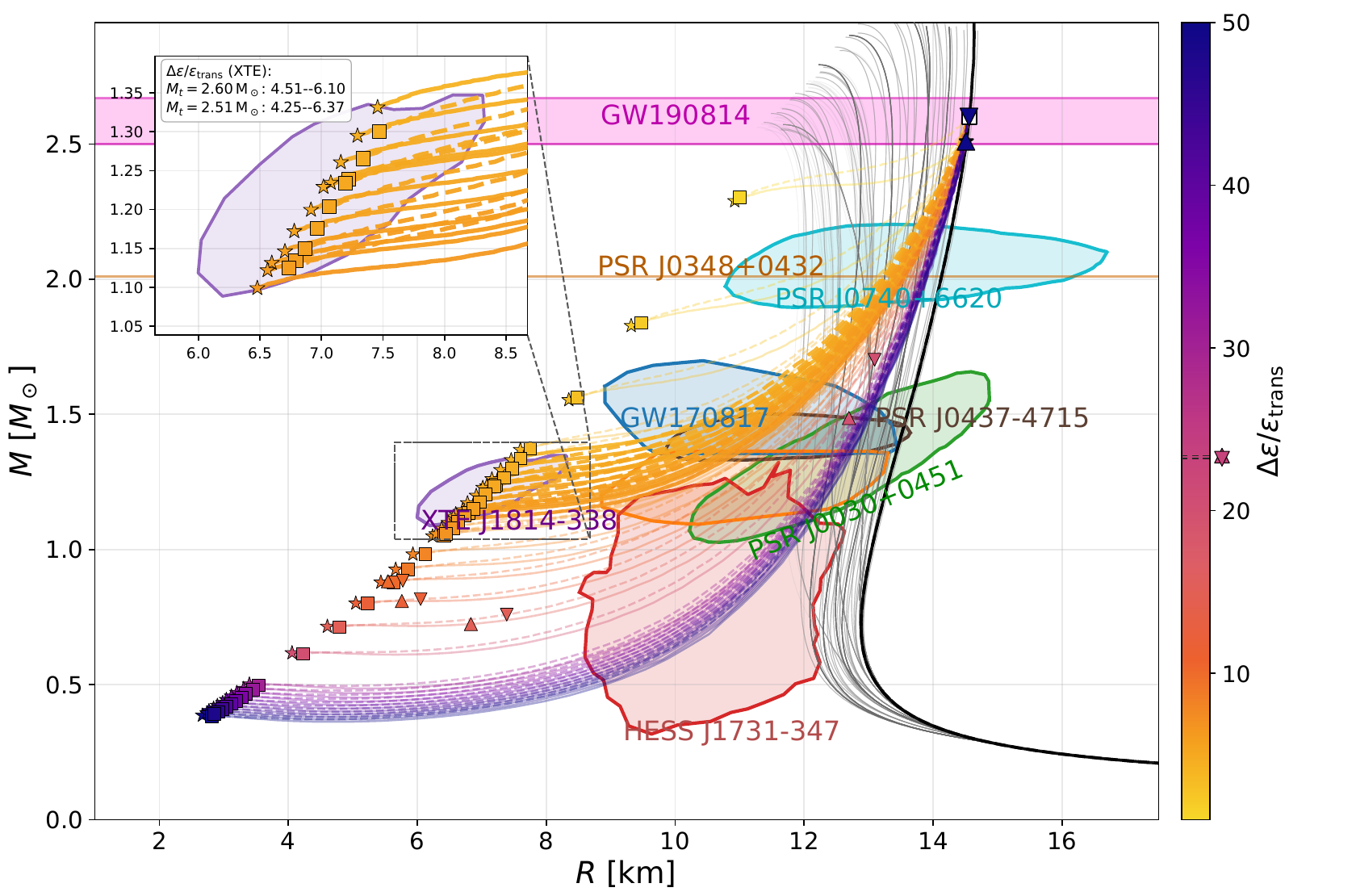}\\[-0.4em]
    \textbf{(b)}
  \end{minipage}
  \caption{$M$--$R$ scans for the BPS+$\chi$EFT(stiff)+PP benchmark, using the
same filters and style conventions as Fig.~\ref{fig:hebeler-xte}.  Panels (a) and (b)  use representative hadronic branches of moderate-$M_{\rm TOV}$ ($2.79\,\Msun$) and highest-$M_{\rm TOV}$ ($ 3.24\,\Msun$),
with corresponding EOS parameters 
$(\Gamma_1,\Gamma_2,\Gamma_3;\rho_{12}/\rho_0,\rho_{23}/\rho_0)
=(1,5.333,8;1.5,3.833)$ and $(3.333,8,8;1.5,6.167)$, respectively.
Triangles and color-bar marks follow Fig.~\ref{fig:hebeler-xte}
and Sec.~\ref{subsec:pqcd}.}
  \label{fig:hebeler-xte-stiff}
\end{figure*}

For the PP benchmark, the parent domain follows Fig.~6 of
Ref.~\cite{Hebeler2013}.  The value $\Gamma_2=0$ is excluded because it
emulates a phase transition inside the hadronic core, whereas the transition is
introduced separately through Eq.~(\ref{eq:css}).  Each four-point combination
is tabulated to $\rho_{\max}=8.3\rho_0$.  If the raw core reaches
$c_s^2=1$, the causal extension in Eq.~(\ref{eq:causal-ext}) is attached before
the maximum-mass cut is applied.  The eight-point pool uses the same ordering
of cuts, with adaptive removal of combinations that violate causality before
their intended PP segment.  The four-point survivors form the dark-gray
background and supply the colored representatives, while the additional
eight-point survivors form the light-gray background in
Figs.~\ref{fig:hebeler-xte} and~\ref{fig:hebeler-xte-stiff}.\footnote{Hybrid
branches are built only from the four-point catalog; the denser eight-point
pool adds pure-hadronic context under the same baseline cuts without a full
hybrid scan on every survivor.}

The independent CS scan forms the full Cartesian product of the values
in Table~\ref{tab:scan-design}, subject only to $a_3\leq2a_2$, before the common
FLT and high-mass filters.  Finally, each displayed retained hadronic branch is
converted into a family of hybrid sequences by placing the Maxwell transition
at $M_t=2.51$ or $2.60\,\Msun$ and scanning
$\Delta\eps/\eps_{\rm trans}$ from 1 to 50.  The main figures use
$c_{s,\rm QM}^2=1$; Sec.~\ref{subsec:qm-stiffness} repeats the PP calculation
with $c_{s,\rm QM}^2=0.8$.

\section{Results}
\label{sec:results}

\subsection{Figure map and the two observational scenarios}
\label{subsec:scenario-map}

Figures~\ref{fig:hebeler-xte} and~\ref{fig:hebeler-xte-stiff} are the
primary mass-radius maps for the soft and stiff BPS+$\chi$EFT matchings,
respectively.  In every panel, the observational contours identify GW170817,
the three NICER pulsars, HESS J1731--347, and XTE J1814--338.  Light-gray curves
are the retained eight-point PP background pool, dark-gray curves are the
retained four-point pool, and the thick black curve is the hadronic sequence
selected for the hybrid scan.  Colored continuations are slow stable hybrid
branches: the color value labels $\Delta\eps/\eps_{\rm trans}$, solid curves
ending in stars use $M_t=2.51\,\Msun$, and dashed curves ending in squares use
$M_t=2.60\,\Msun$.
Filled triangles mark the first point along each
slow stable branch at which the central density reaches
$n_{\rm pQCD}=40n_0$ (upright for $M_t=2.51\,\Msun$, inverted for
$M_t=2.60\,\Msun$); if that density is already attained immediately after the
Maxwell jump, the triangle is stacked on the transition symbol.  Dashed marks
on the color bar give the corresponding critical $\Delta\eps/\eps_{\rm trans}$.
These annotations are defined in Sec.~\ref{subsec:pqcd} and are repeated on the
tidal sequences of Figs.~\ref{fig:app-hebeler-tidal-soft}
and~\ref{fig:app-hebeler-tidal-stiff}.
Each inset magnifies the XTE contour and shows only the
density-jump interval that intersects it.  Panels (a) and (b) then isolate,
respectively, a moderate-$M_{\rm TOV}$ representative and the
highest-$M_{\rm TOV}$ representative.  This visual hierarchy is used
consistently in the PP, CS, and tidal figures below and in the sound-speed
figures in Appendix~\ref{app:sound-speed-profiles}.

The main results can be categorized into two possible scenarios, with the shared feature that the pure hadronic branch retains enough high-density support to cover the GW190814-scale mass:
in scenario 1 [e.g., Figs.~\ref{fig:hebeler-xte}(a) and~\ref{fig:hebeler-xte-stiff}(a)], this hadronic branch also passes through the HESS J1731--347 region, with the slow stable hybrid
continuation generated by the phase transition supporting the XTE J1814--338 small-radius crossing.
In scenario 2 [e.g., Figs.~\ref{fig:hebeler-xte}(b) and~\ref{fig:hebeler-xte-stiff}(b)], the HESS J1731--347 region is missed by most hadronic branches,
yet can instead be reached by the slow stable branch with very large $\Delta \eps/\eps_{\rm trans}$. 
Note that scenario 1 is the case where all constraints are met by a single EOS construction, while scenario 2 gives the merit that either XTE J1814--338 or HESS J1731--347 are explained by the slow stable branch in different $\Delta \eps/\eps_{\rm trans}$ regime. 

This required high-density support is reflected in the sound speed along the hadronic branch.  As shown in Appendix~\ref{app:sound-speed-profiles}, the transition configurations relevant for the viable solutions occur where the central sound speed has already grown large and, in most benchmark cases, approaches the causal limit near the high-mass transition point.  This does not contradict conformality-motivated studies: the conformal value is not a robust upper bound at neutron-star densities, as emphasized by Refs.~\cite{Tews2018, Bedaque2015}, while pQCD-driven high-density softening enters conformality inferences through additional asymptotic-QCD assumptions; Appendix~\ref{app:sound-speed-profiles} gives a more detailed discussion.

Considering that scenario 1 maps to relatively soft EOSs in the intermediate densities, the GW170817 constraint is easily met by the hadronic branch in scenario 1.  In scenario 2, the GW170817 compatibility can still be saved by the slow stable branch when the hadronic branch is too stiff in the corresponding region. These are explicitly demonstrated by the shaded GW170817 $M$-$R$ region in Figs.~\ref{fig:hebeler-xte} and \ref{fig:hebeler-xte-stiff}, and the tidal deformability results in Figs.~\ref{fig:app-hebeler-tidal-soft}, \ref{fig:app-hebeler-tidal-stiff}.  The tidal response follows the same two-stage structure as
the mass-radius result.  The moderate-$M_{\rm TOV}$ hadronic branches remain below the reference
$\Lambda_{1.4\,\Msun}=800$ line, whereas the highest-$M_{\rm TOV}$ hadronic branches lie close to
or slightly above it.  After the Maxwell jump, however, the slow stable hybrid branches that enter
the XTE J1814--338 region have much smaller tidal deformabilities, typically $\Lambda=\mathcal{O}(10)$ at the
XTE masses.  The same $n_c=n_{\rm pQCD}$ triangles appear on
Figs.~\ref{fig:app-hebeler-tidal-soft} and~\ref{fig:app-hebeler-tidal-stiff},
stacked on the transition markers whenever the jump already overshoots
$40n_0$.  Among the hybrid sequences in those figures that satisfy
$\Lambda_{1.4\,\Msun}<800$, some of the large-jump members that still exist
at $1.4\,\Msun$ already have $n_c>n_{\rm pQCD}$ at that mass.  The
XTE-intersecting branches still satisfy
$\Lambda_{1.4\,\Msun}<800$ and $n_c<n_{\rm pQCD}=40n_0$ at
$M=1.4\,\Msun$.  Thus, the branch that solves the XTE small-radius problem is not in tension with the
usual GW170817 upper-bound direction.  Note that we do not impose lower bounds or stricter upper bounds on tidal deformability in some later refinements after the original GW170817 analyses.  This is because such refinements often rely
on quasiuniversal relations, or kilonova/postmerger arguments that did not include strong phase transitions and the resulting slow stable hybrid branches.  \par\onecolumngrid
\begingroup
\setlength{\intextsep}{10pt}
\begin{figure}[!h]
  \hsize=\textwidth
  \linewidth=\textwidth
  \centering
  \begin{minipage}[t]{0.49\textwidth}
    \centering
    \includegraphics[width=\linewidth]{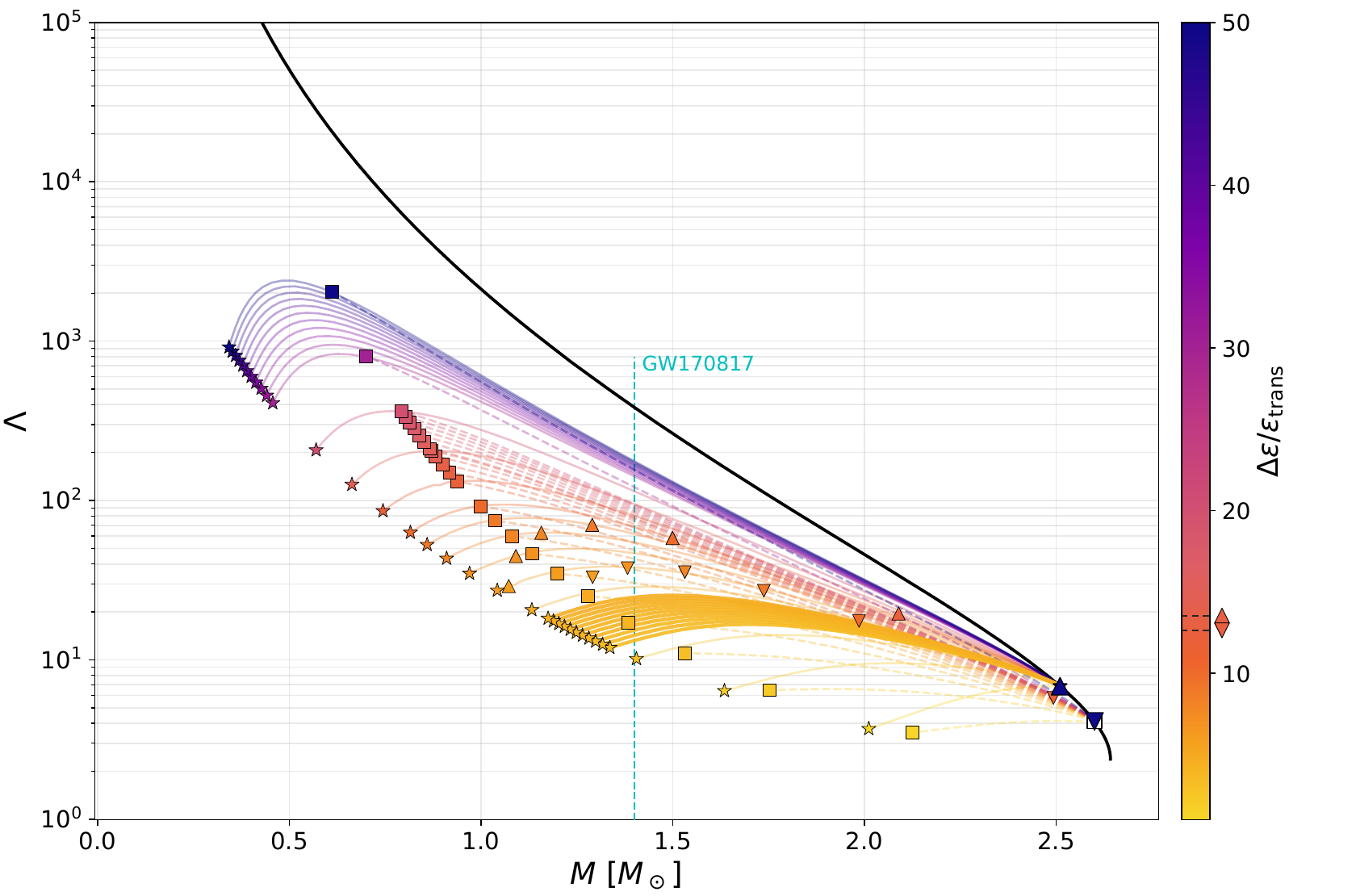}\\[-0.4em]
    \textbf{(a)}
  \end{minipage}\hfill
  \begin{minipage}[t]{0.49\textwidth}
    \centering
    \includegraphics[width=\linewidth]{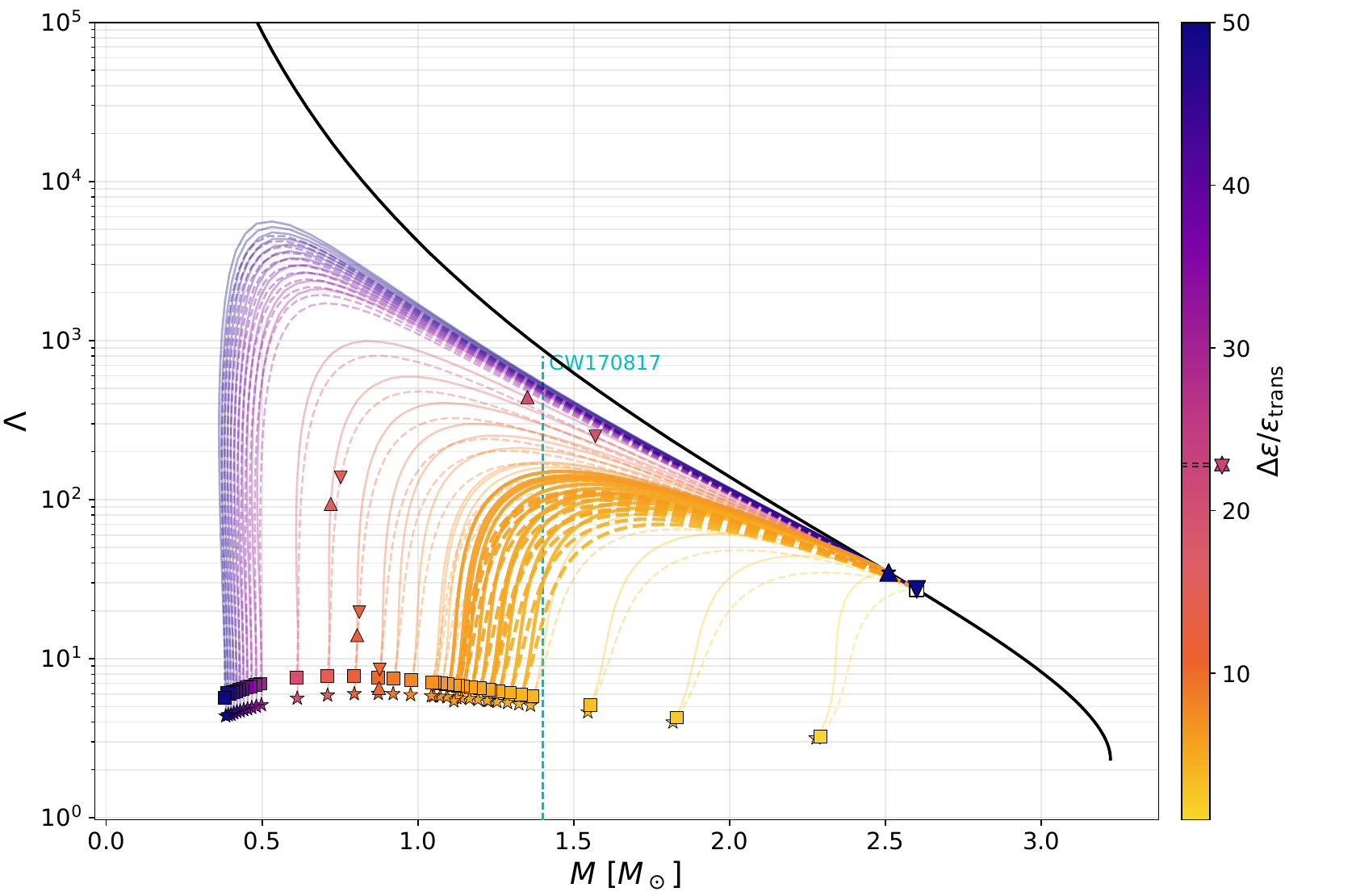}\\[-0.4em]
    \textbf{(b)}
  \end{minipage}
  \caption{Tidal deformability $\Lambda(M)$ for the benchmark scans
  of Fig.~\ref{fig:hebeler-xte}, using the same style conventions.
  The $n_c=n_{\rm pQCD}$ triangles and color-bar ticks follow
  Fig.~\ref{fig:hebeler-xte}.
  The cyan
  dashed segment marks $\Lambda_{1.4\,\Msun}<800$ (GW170817~\cite{Abbott2017GW170817}).}
  \label{fig:app-hebeler-tidal-soft}
\end{figure}
\FloatBarrier
\endgroup
\begin{figure*}[t]
  \hsize=\textwidth
  \linewidth=\textwidth
  \centering
  \begin{minipage}[t]{0.49\textwidth}
    \centering
    \includegraphics[width=\linewidth]{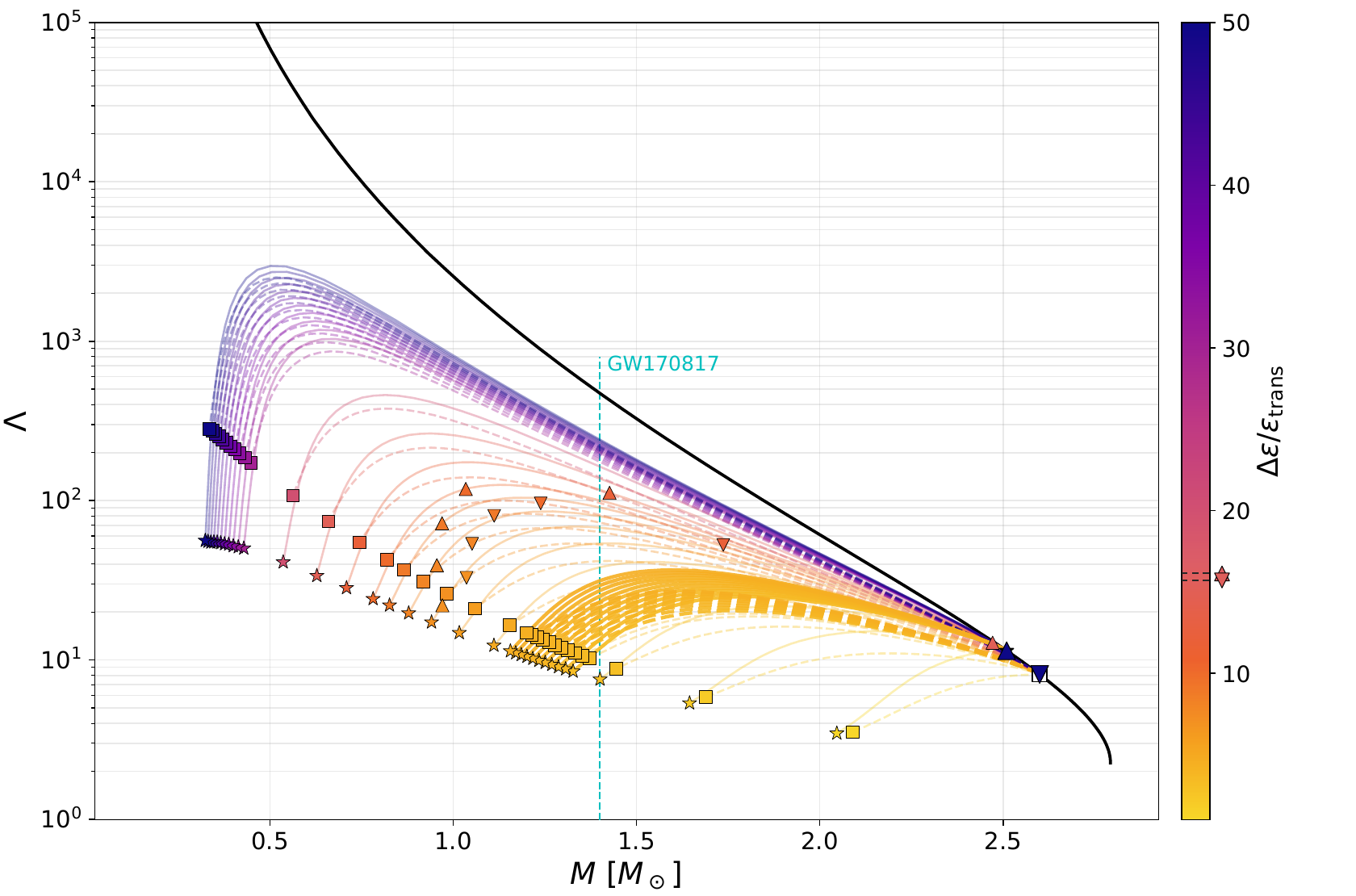}\\[-0.4em]
    \textbf{(a)}
  \end{minipage}\hfill
  \begin{minipage}[t]{0.49\textwidth}
    \centering
    \includegraphics[width=\linewidth]{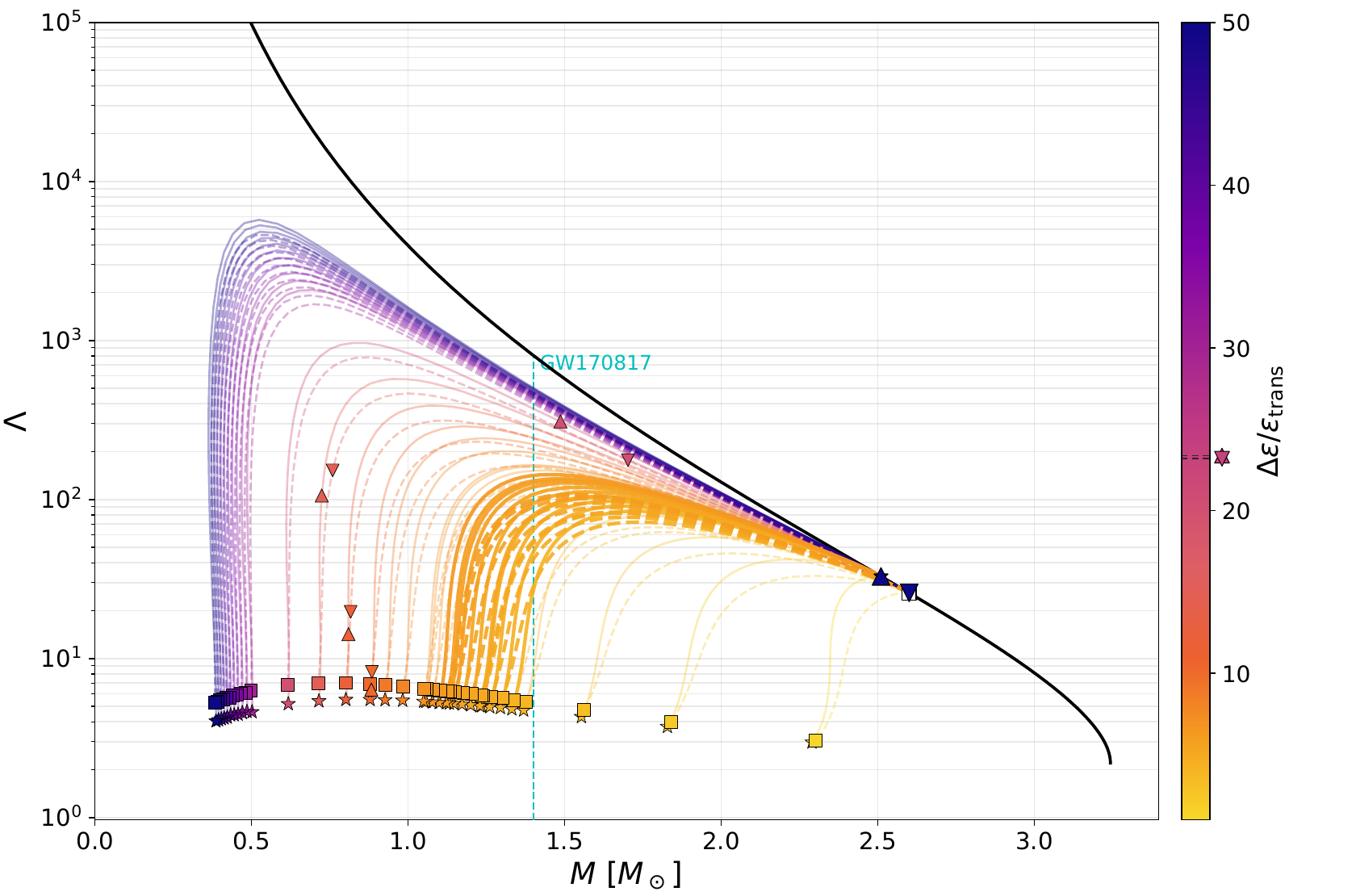}\\[-0.4em]
    \textbf{(b)}
  \end{minipage}
  \caption{Tidal deformability $\Lambda(M)$ for the benchmark scans
  of Fig.~\ref{fig:hebeler-xte-stiff}, using the same style conventions.
  The $n_c=n_{\rm pQCD}$ triangles and color-bar ticks follow
  Fig.~\ref{fig:hebeler-xte}.
  The cyan dashed
  segment marks $\Lambda_{1.4\,\Msun}<800$ (GW170817~\cite{Abbott2017GW170817}).}
  \label{fig:app-hebeler-tidal-stiff}
\end{figure*}

\newpage
\twocolumngrid

Particularly, it is known that strong first-order phase transitions can violate the universal-relation assumptions used in some of these inferences~\cite{han2019tidal,hoyos2022holographic}.

\subsection{Effects of transition mass and strength}
\label{subsec:transition-effects}

The XTE J1814--338 crossing is controlled primarily by transition placement and jump strength.  For
$\csssq=1$, transitions near $M_t\simeq2.5$--$2.6\,\Msun$ yield intersections with XTE J1814--338 for
density jumps of order a few times $\eps_{\rm trans}$ in the polytropic benchmark, as illustrated in
Figs.~\ref{fig:hebeler-xte} and \ref{fig:hebeler-xte-stiff}.  The CS-parametrization comparison in
Figs.~\ref{fig:greif-soft-xte} and~\ref{fig:greif-stiff-xte} gives comparable windows.  The finite width of this crossing window is itself a diagnostic feature of the mechanism.

First, we can see that an earlier transition (i.e., a smaller $M_t$) starts the
hybrid continuation closer to the low-mass end and leaves a longer stable segment over which the
sequence can bend toward small masses and radii.

As for the transition strength, we see that a smaller jump leaves the
hybrid branch too close to the hadronic sequence and therefore too large in radius; a larger jump
bends the branch too abruptly, missing the XTE J1814--338 region from the compact side or leaving
the relevant mass range.  The crossing is therefore a diagnostic of a strong transition in a particular placement range. Across the displayed scans, the
XTE-compatible jumps are still too small to pass through the HESS J1731--347 central region, whereas much
larger jumps, roughly $\deltaratio\simeq10$--$50$, can drive a compact slow stable branch deep through the
HESS J1731--347 region.
These large jumps are also supercritical in the Seidov sense:
for a Maxwell transition, the onset of an
infinitesimal high-density core destabilizes the equilibrium sequence under
the usual rapid-conversion treatment when the jump exceeds the Seidov
threshold~\cite{Seidov1971,Alford2013CSS},
\begin{equation}
  \left(\frac{\Delta\eps}{\eps_{\rm trans}}\right)_{\rm Seidov}
  =\frac{1}{2}+\frac{3}{2}
  \frac{P_{\rm trans}}{\eps_{\rm trans}}.
  \label{eq:seidov}
\end{equation}
For the eight representative PP transition points in
Figs.~\ref{fig:hebeler-xte} and~\ref{fig:hebeler-xte-stiff}, the right-hand
side ranges from $0.95$ to $1.34$.  Thus the jumps of a few required by XTE,
and especially the $10$--$50$ jumps that approach HESS, are strongly
supercritical in the Seidov sense.  This explains why the corresponding
posttransition segment would normally be discarded in a rapid-conversion
analysis and makes the conversion type essential to our mechanism.


\begin{figure*}[t]
  \centering
  \begin{minipage}[t]{0.49\textwidth}
    \centering
    \includegraphics[width=\linewidth]{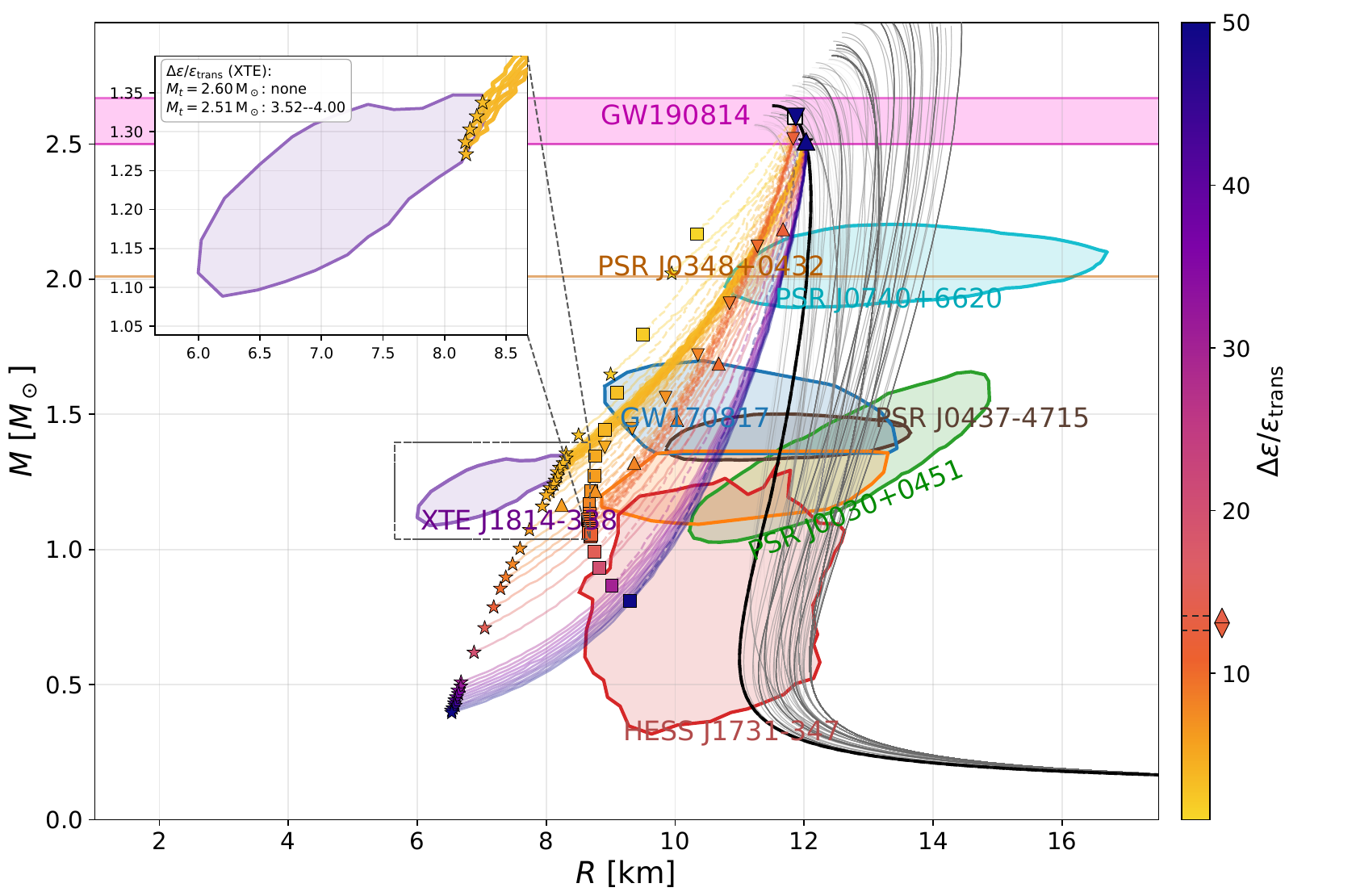}\\[-0.4em]
    \textbf{(a)}
  \end{minipage}\hfill
  \begin{minipage}[t]{0.49\textwidth}
    \centering
    \includegraphics[width=\linewidth]{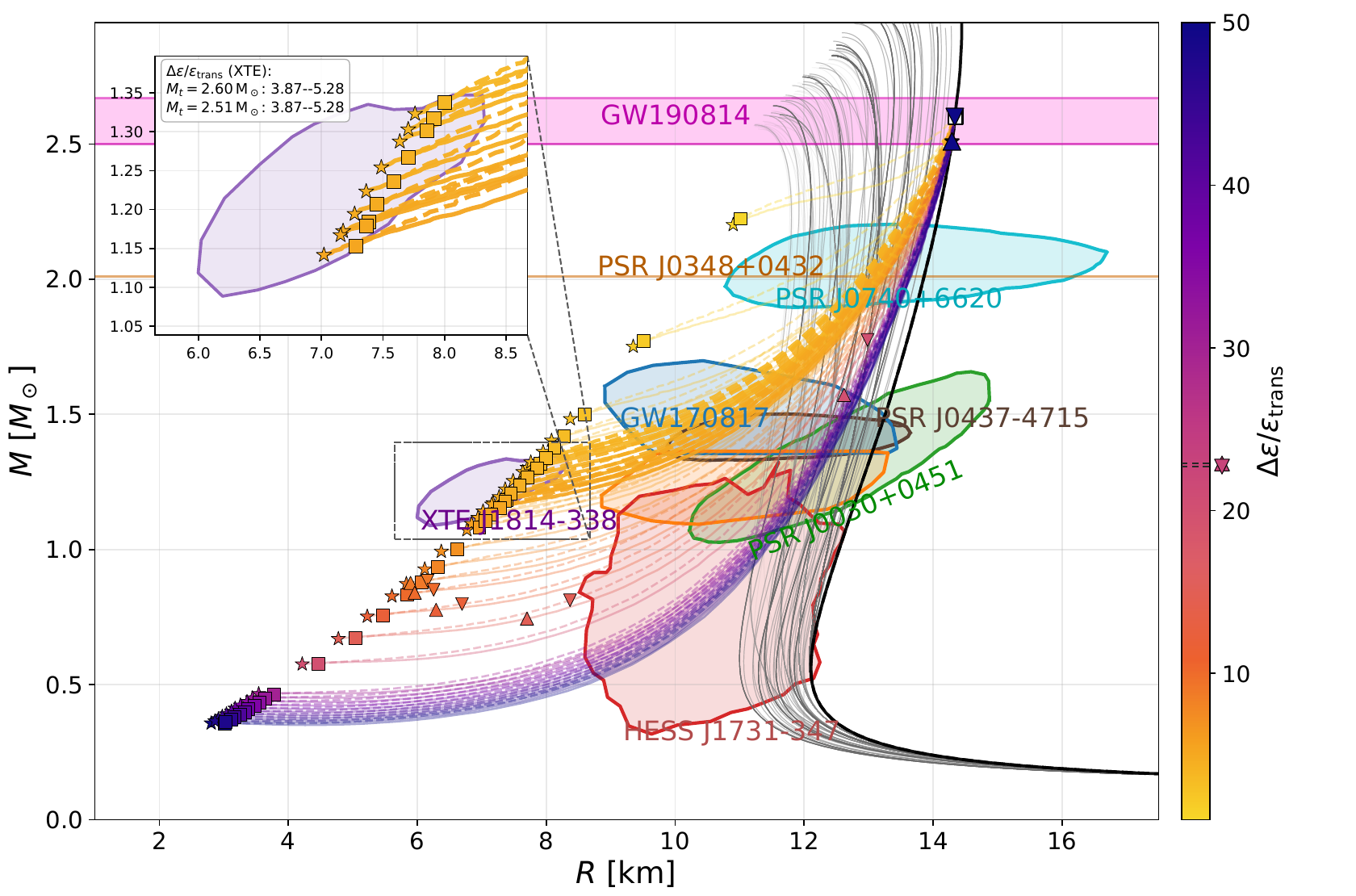}\\[-0.4em]
    \textbf{(b)}
  \end{minipage}
  \caption{The same case as Fig.~\ref{fig:hebeler-xte} but with $\csssq=0.8$, using the same color and symbol conventions.  The black hadronic sequences in panels (a,b) again use
$(\Gamma_1,\Gamma_2,\Gamma_3;\rho_{12}/\rho_0,\rho_{23}/\rho_0)
=(1,5.333,0.5;1.5,8.5)$ and $(4.5,8,0.5;1.5,6.167)$, respectively.  The reduced quark-matter stiffness shortens the slow stable branches relative to the $\csssq=1$ benchmark.
Triangles and color-bar marks follow Fig.~\ref{fig:hebeler-xte}.}
  \label{fig:app-hebeler-xte-cs080-soft}
\end{figure*}

\subsection{Effects of posttransition stiffness}
\label{subsec:qm-stiffness}

\begin{figure*}[t]
  \centering
  \begin{minipage}[t]{0.49\textwidth}
    \centering
    \includegraphics[width=\linewidth]{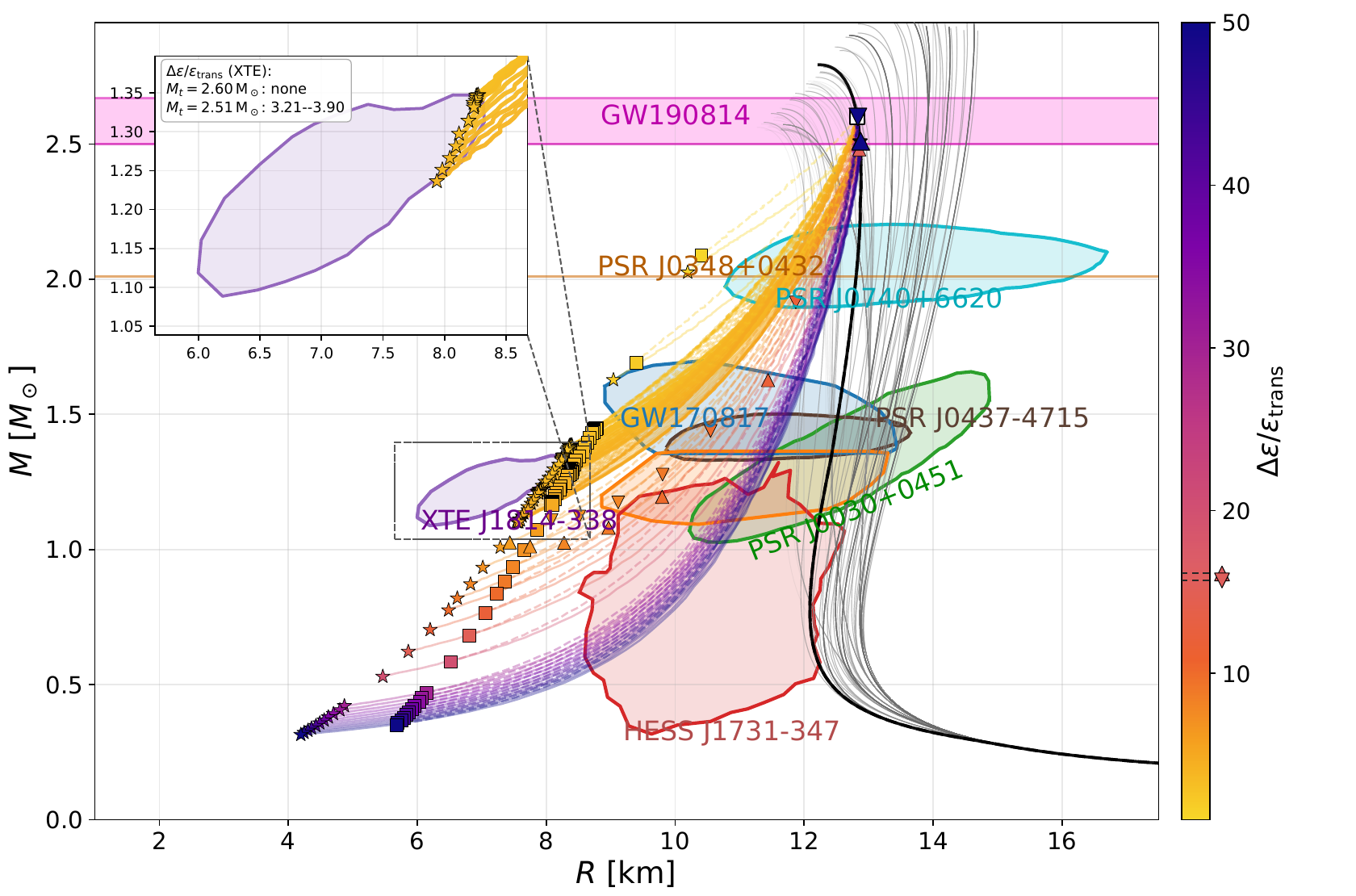}\\[-0.4em]
    \textbf{(a)}
  \end{minipage}\hfill
  \begin{minipage}[t]{0.49\textwidth}
    \centering
    \includegraphics[width=\linewidth]{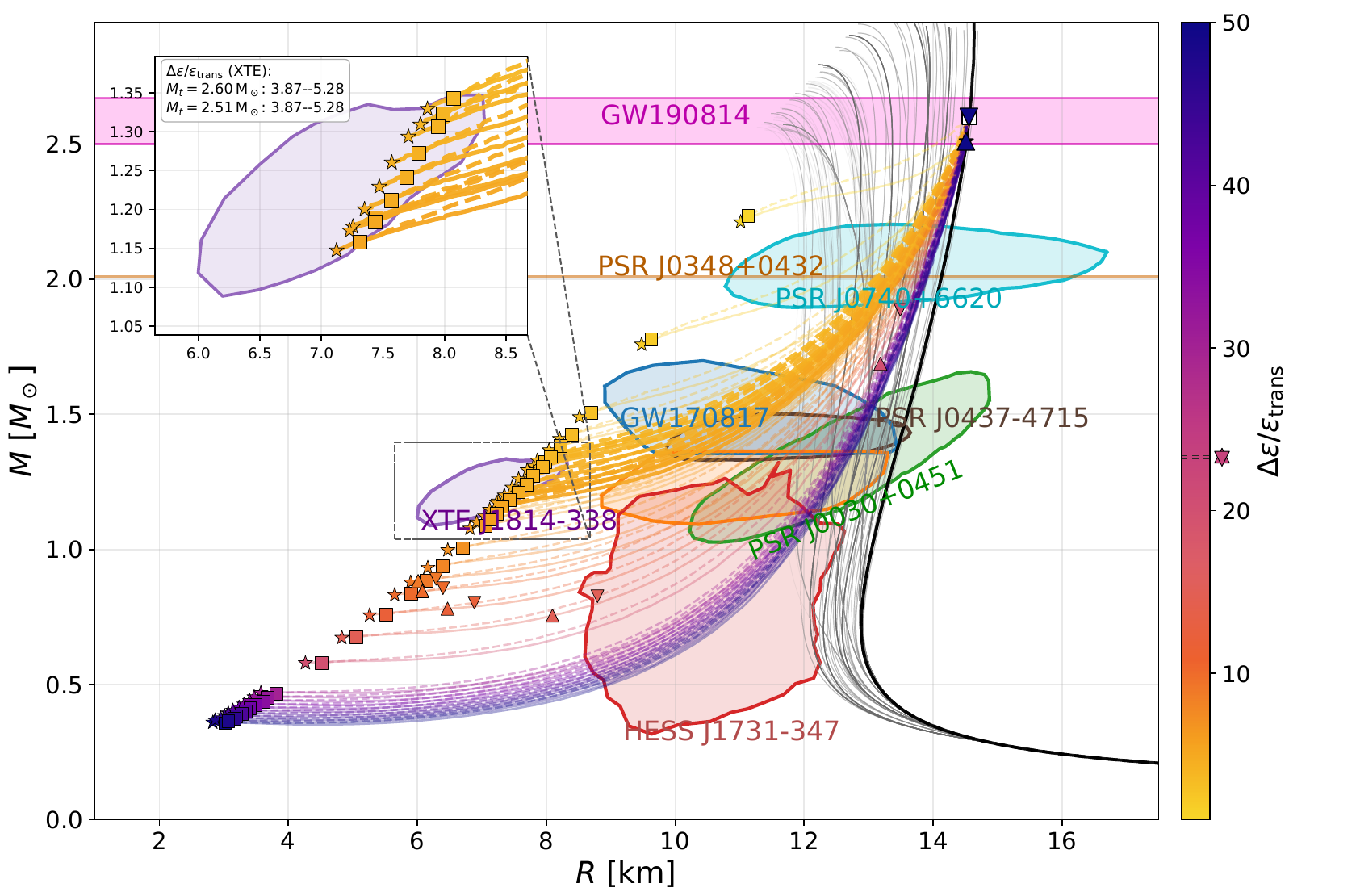}\\[-0.4em]
    \textbf{(b)}
  \end{minipage}
  \caption{
  The same case as Fig.~\ref{fig:hebeler-xte-stiff} but with $\csssq=0.8$, using the same color and symbol conventions.  The black hadronic sequences in panels (a,b) again use
$(\Gamma_1,\Gamma_2,\Gamma_3;\rho_{12}/\rho_0,\rho_{23}/\rho_0)
=(1,5.333,8;1.5,3.833)$ and $(3.333,8,8;1.5,6.167)$, respectively.  The reduced quark-matter stiffness shortens the slow stable branches relative to the $\csssq=1$ benchmark.
Triangles and color-bar marks follow Fig.~\ref{fig:hebeler-xte}.}
  \label{fig:app-hebeler-xte-cs080-stiff}
\end{figure*}

To quantify this posttransition stiffness dependence, we repeat the
polytropic benchmark scans of Figs.~\ref{fig:hebeler-xte}
and~\ref{fig:hebeler-xte-stiff} with $\csssq=0.8$ instead of the main-scan
choice $\csssq=1$, while keeping
the same hadronic sequences, transition masses $M_t=2.51$ and $2.60\,\Msun$, and
density-jump scan.  Softening the posttransition phase shortens the slow stable continuation after
the Maxwell jump and increases the typical radii of the slow stable branches.  The branch therefore
has less leverage to move toward the low-mass, small-radius region required by XTE J1814--338.  This
effect acts in the same direction as the aforementioned increase in transition mass, matching the analogous
behavior of the complete $c_s^2$ scan of self-bound hybrid
stars~\cite{Zhang2026SelfBound}.  For the examples with highest-$M_{\rm TOV}$ hadronic branches
[Figs.~\ref{fig:app-hebeler-xte-cs080-soft}(b) and~\ref{fig:app-hebeler-xte-cs080-stiff}(b)], the
number of XTE-intersecting branches remains nine at each transition mass, with a shifted $\deltaratio$ window
relative to the $\csssq=1$ case.  For the moderate-$M_{\rm TOV}$ soft hadronic branch
[Fig.~\ref{fig:app-hebeler-xte-cs080-soft}(a)], intersections with XTE J1814--338 persist only at $M_t=2.51\,\Msun$ (five of
thirty-four scanned ratios), while none remain at $M_t=2.60\,\Msun$.  For the moderate-$M_{\rm TOV}$
stiff hadronic branch [Fig.~\ref{fig:app-hebeler-xte-cs080-stiff}(a)], intersections with XTE J1814--338 persist at
$M_t=2.51\,\Msun$ (fourteen of eighty-two scanned ratios), while none remain at
$M_t=2.60\,\Msun$.

\subsection{Central densities and the pQCD domain}
\label{subsec:pqcd}

To quantify the high-density extrapolation, we reconstruct the baryon
density of the cold CSS phase from $d\eps=(\eps+P)\,dn/n$ and impose
chemical-potential continuity across the Maxwell jump.  If $n_-$ denotes the
hadronic density immediately below the transition and
$\eps_+=\eps_{\rm trans}+\Delta\eps$, this gives
\begin{equation}
  \begin{aligned}
  &n_+=n_-\frac{\eps_++P_{\rm trans}}
                    {\eps_{\rm trans}+P_{\rm trans}},\\
  &n(P)=n_+
  \left[\frac{\eps(P)+P}{\eps_++P_{\rm trans}}\right]^{1/(1+\csssq)}.
  \end{aligned}
  \label{eq:css-number-density}
\end{equation}
For all XTE-intersecting points of the four representative hadronic
branches in Figs.~\ref{fig:hebeler-xte} and~\ref{fig:hebeler-xte-stiff}
(the solid black curves) and both transition masses,
Eq.~\eqref{eq:css-number-density} gives
$15.6\lesssim n_c/n_0\lesssim28.8$, with
$3.64\lesssim\eps_c\lesssim8.53~{\rm GeV\,fm^{-3}}$.  At the deliberately
extreme $\Delta\eps/\eps_{\rm trans}=50$ marginal end points, the ranges become
$186\lesssim n_c/n_0\lesssim324$ and
$52.7\lesssim\eps_c\lesssim127~{\rm GeV\,fm^{-3}}$.  These values were
evaluated directly from the archived branch files; the latter range is driven
mainly by the enormous Maxwell discontinuity.

The same reconstruction is used to annotate the hybrid sequences.
We adopt $n_{\rm pQCD}=40n_0$ as a conventional matching density at which
current EOS constructions begin to consult perturbative
QCD~\cite{Gorda2023QCDInput,Annala:2023cwx}.
Importantly, the approach to this scale is a posttransition hybrid effect:
across the pure-hadronic pools of
Figs.~\ref{fig:app-hebeler-cs2-vs-pc} and~\ref{fig:app-greif-cs2-vs-pc},
every sequence that reaches $M_t=2.51$ or $2.60\,\Msun$ still has
$n_c\lesssim5n_0$ at that mass (hence $n_-<n_{\rm pQCD}$ for all
Maxwell jumps considered here), so the entire hadronic segment
$M\le M_t$ remains well below $n_{\rm pQCD}$.
On each slow stable branch, a
filled triangle marks the first stellar configuration with
$n_c=n_{\rm pQCD}$: upright triangles for $M_t=2.51\,\Msun$ and inverted
triangles for $M_t=2.60\,\Msun$.  If the quark core already exceeds
$40n_0$ immediately after the Maxwell jump ($n_+\ge n_{\rm pQCD}$), the
triangle is stacked on the transition symbol rather than placed later along
the hybrid continuation.  The identical convention is used on the
$\Lambda(M)$ sequences of Figs.~\ref{fig:app-hebeler-tidal-soft}
and~\ref{fig:app-hebeler-tidal-stiff}.

Setting $n_+=n_{\rm pQCD}$ in Eq.~\eqref{eq:css-number-density} gives
the critical jump at which this stacking occurs,
\begin{equation}
  \left(\frac{\Delta\eps}{\eps_{\rm trans}}\right)_{\rm crit}
  =\frac{n_{\rm pQCD}}{n_-}
  \left(1+\frac{P_{\rm trans}}{\eps_{\rm trans}}\right)
  -1-\frac{P_{\rm trans}}{\eps_{\rm trans}}
  \label{eq:npqcd-crit}
\end{equation}
Dashed ticks on the color bars of
Figs.~\ref{fig:hebeler-xte}--\ref{fig:greif-stiff-xte} mark this threshold
for the two transition masses, with the same upright/inverted triangle
as on the sequences.  For these four panels the critical jumps are
$\bigl(\Delta\eps/\eps_{\rm trans}\bigr)_{\rm crit}=13.6$ and $12.6$ in
Fig.~\ref{fig:hebeler-xte}(a), $22.9$ and $22.7$ in
Fig.~\ref{fig:hebeler-xte}(b), $16.1$ and $15.7$ in
Fig.~\ref{fig:hebeler-xte-stiff}(a), and $23.4$ and $23.2$ in
Fig.~\ref{fig:hebeler-xte-stiff}(b), quoting $M_t=2.51\,\Msun$ then
$2.60\,\Msun$.  Because $n_+$ is fixed by the Maxwell matching and does not
involve $\csssq$, $\bigl(\Delta\eps/\eps_{\rm trans}\bigr)_{\rm crit}$ is
unchanged between the
$\csssq=1$ and $\csssq=0.8$ figures; only an interior crossing along a
given branch moves when the quark-matter stiffness is reduced.  The CS
panels of Figs.~\ref{fig:greif-soft-xte} and~\ref{fig:greif-stiff-xte}
yield a similar critical-jump window,
$\bigl(\Delta\eps/\eps_{\rm trans}\bigr)_{\rm crit}\simeq21$--$24$.

In this setting, pQCD is not automatically far above the densities
reached by the slow stable stars.  Even the XTE-compatible configurations
enter the tens-of-$n_0$ regime, and the largest-jump end points lie far beyond
densities at which high-density QCD information must be
consulted~\cite{Gorda2023QCDInput,Annala:2023cwx,Mariani:2025nud}.
Equation~\eqref{eq:npqcd-crit} makes the split visible on the figures:
XTE-compatible jumps remain below the color-bar threshold, so any
$n_c=n_{\rm pQCD}$ mark on those branches lies later along the hybrid
continuation.  For scenario~2 sequences that actually cross the
HESS J1731--347 region, the milder HESS-crossing jumps still have
$n_c<n_{\rm pQCD}$ at that region, whereas the larger jumps already sit
above $40n_0$ there and often already at onset, which is why their
triangles pile up on the $M_t$ symbols.
The $\Lambda(M)$ sequences of
Figs.~\ref{fig:app-hebeler-tidal-soft} and~\ref{fig:app-hebeler-tidal-stiff}
show the same split at $M=1.4\,\Msun$: among the hybrid branches with
$\Lambda_{1.4\,\Msun}<800$, the large-jump examples already have
$n_c>n_{\rm pQCD}$, while the XTE-intersecting ones remain below
$40n_0$.
Those ultradense end points are therefore properties of an
unmatched CSS extrapolation, not quantitative predictions for physical
stars.  A future construction should match the quark phase to pQCD and
recompute both equilibrium and conversion dynamics; until then, the
largest-jump part of scenario 2 is illustrative, whereas the more moderate
XTE window is the less extreme result.

Comparing Fig.~\ref{fig:hebeler-xte}  and Fig.~\ref{fig:hebeler-xte-stiff}, we can also see that a softer crust/outer core can help the hadronic branch delve deeper into the HESS J1731--347 region, matching the expectation that a softer EOS yields smaller radii.

\par\onecolumngrid
\begingroup
\setlength{\intextsep}{10pt}
\begin{figure}[!h]
  \hsize=\textwidth
  \linewidth=\textwidth
  \centering
  \begin{minipage}[t]{0.49\textwidth}
    \centering
    \includegraphics[width=\linewidth]{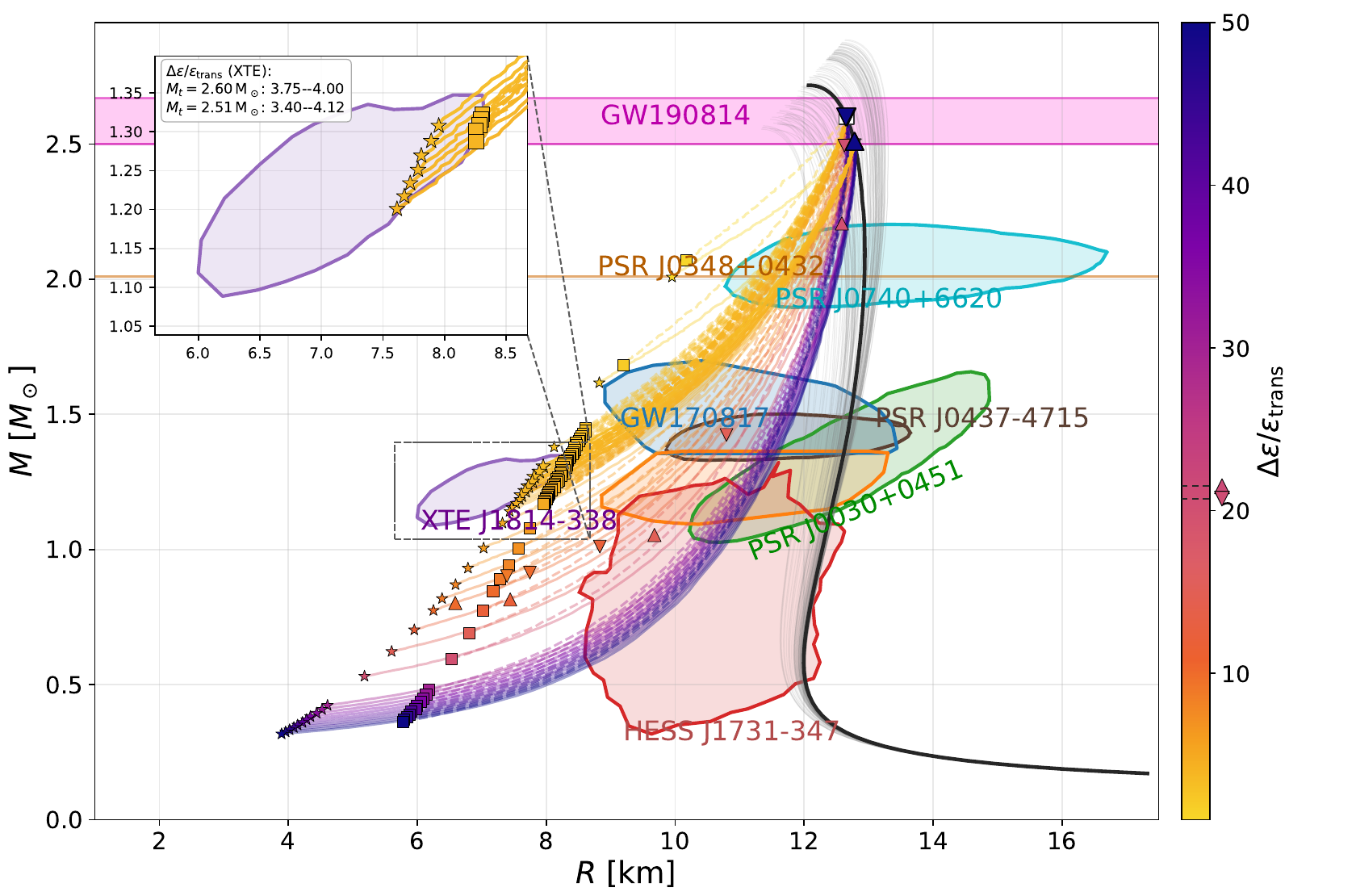}\\[-0.4em]
    \textbf{(a)}
  \end{minipage}\hfill
  \begin{minipage}[t]{0.49\textwidth}
    \centering
    \includegraphics[width=\linewidth]{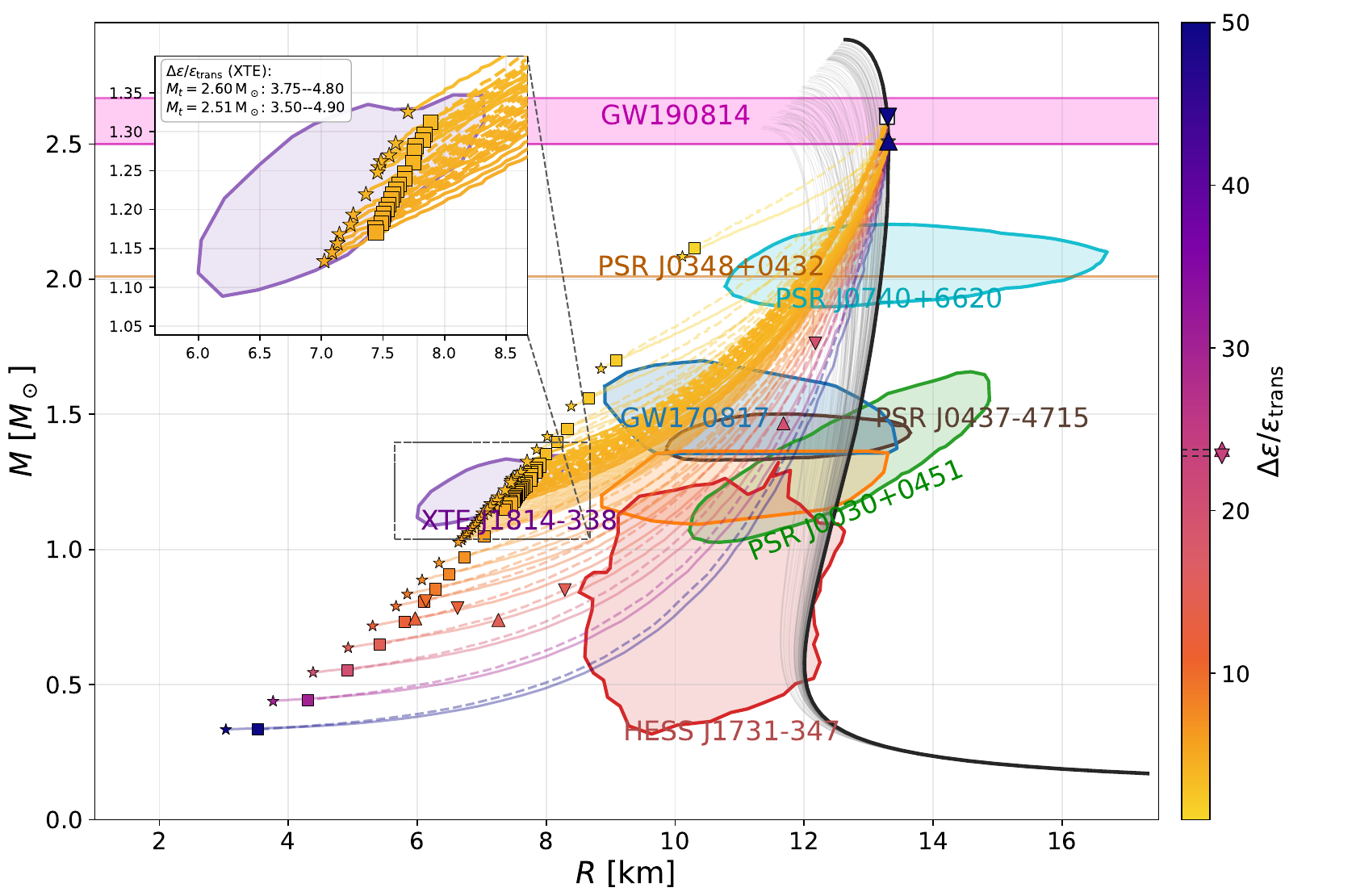}\\[-0.4em]
    \textbf{(b)}
  \end{minipage}
  \caption{$M$--$R$ scans for the BPS+$\chi$EFT(soft)+CS construction, using the
same filters and style conventions as Fig.~\ref{fig:hebeler-xte}.  Panels
(a) and (b) use representative hadronic  branches of moderate-$M_{\rm TOV}$ ($2.717\,\Msun$) and highest-$M_{\rm TOV}$ ($2.886\,\Msun$), with corresponding EOS parameters
$(a_1,a_2,a_3,a_4,a_5,a_6,a_7)=(1.8,12,6,3,0.7,-0.668,1.0)$ and
$(1.5,4,8,3,1.4,-1.544,1.0)$, respectively.
Triangles and color-bar marks follow Fig.~\ref{fig:hebeler-xte}.}
  \label{fig:greif-soft-xte}
\end{figure}
\FloatBarrier
\endgroup
\begin{figure*}[t]
  \hsize=\textwidth
  \linewidth=\textwidth
  \centering
  \begin{minipage}[t]{0.49\textwidth}
    \centering
    \includegraphics[width=\linewidth]{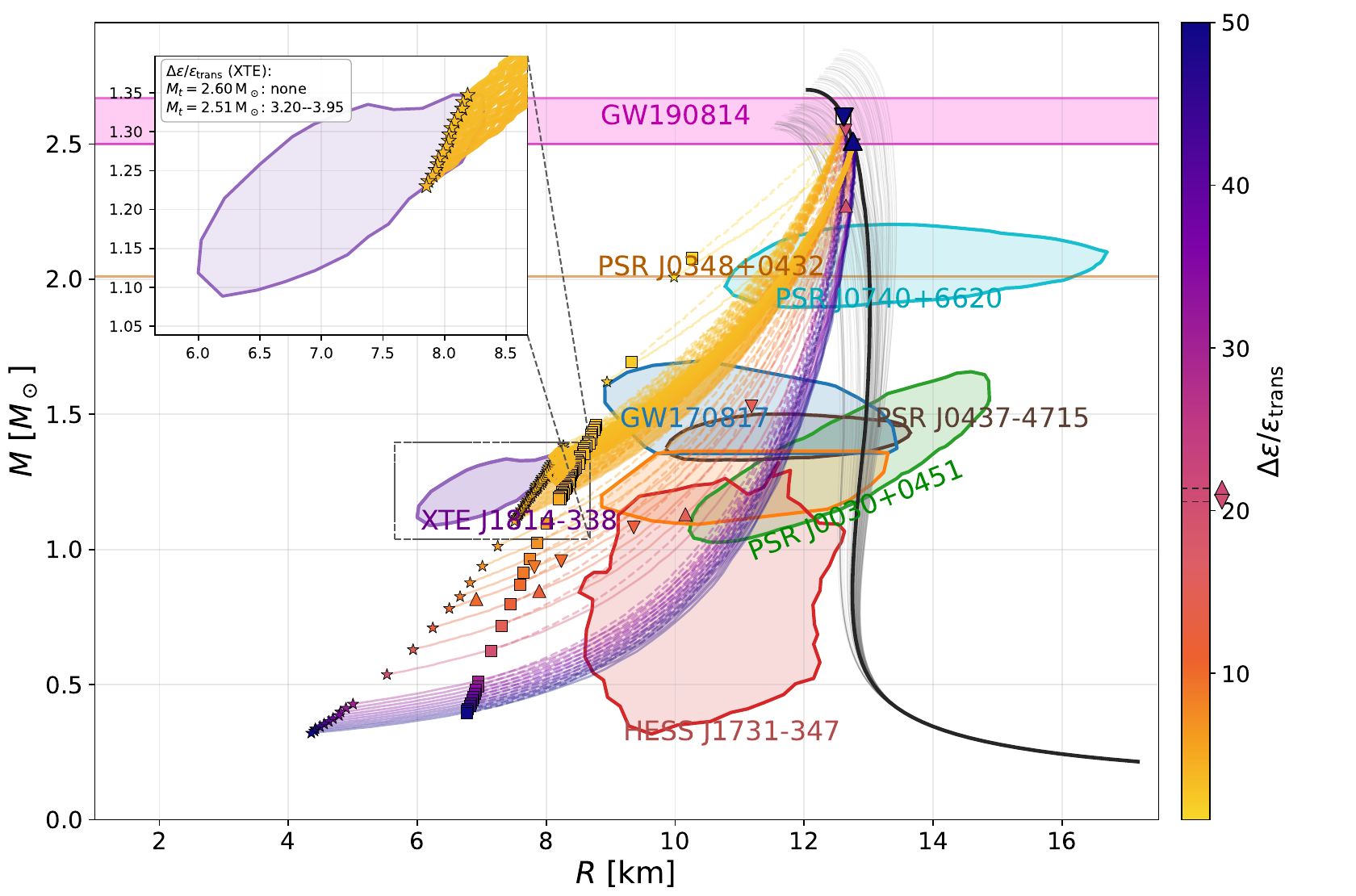}\\[-0.4em]
    \textbf{(a)}
  \end{minipage}\hfill
  \begin{minipage}[t]{0.49\textwidth}
    \centering
    \includegraphics[width=\linewidth]{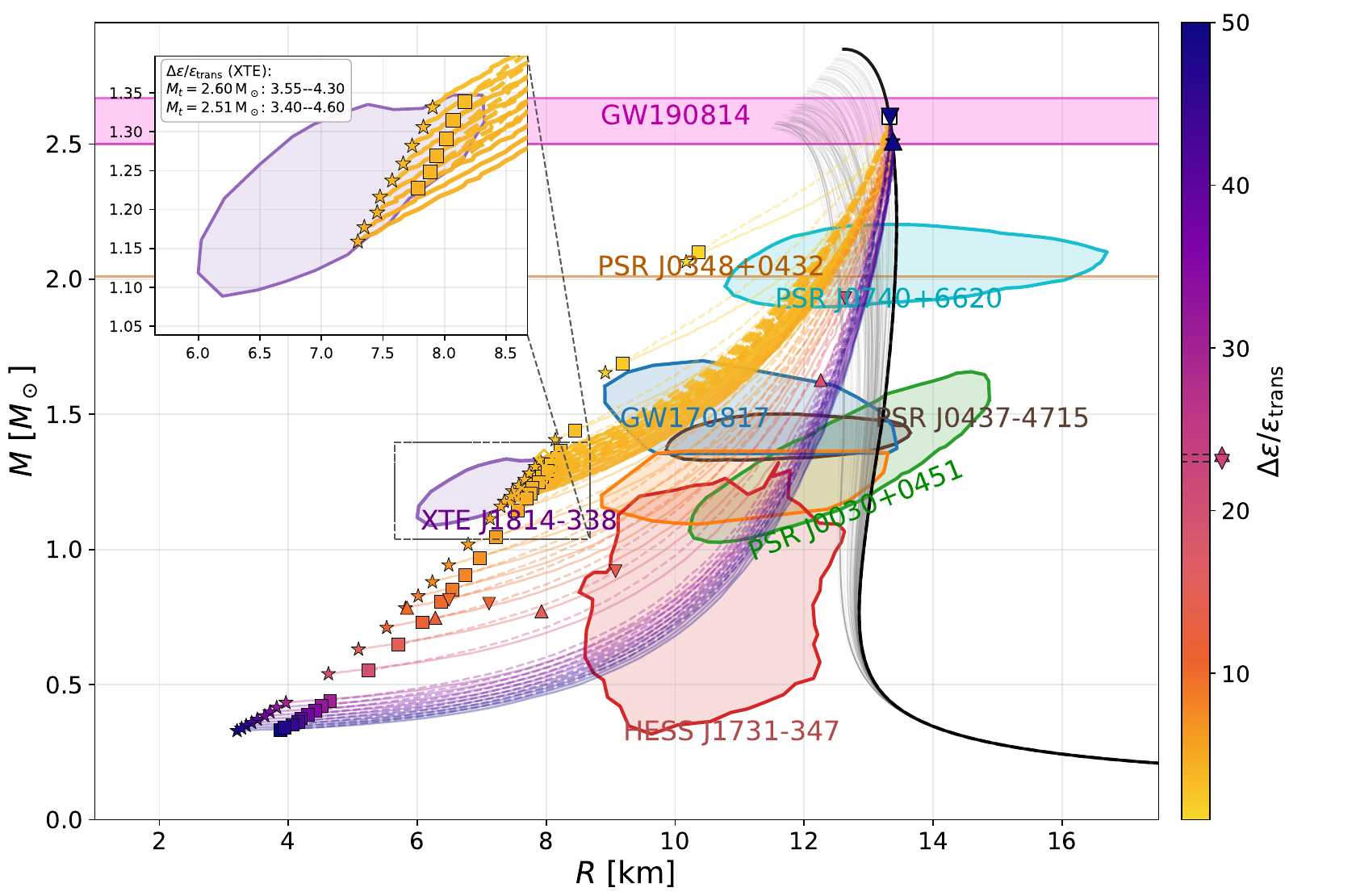}\\[-0.4em]
    \textbf{(b)}
  \end{minipage}
  \caption{Same as Fig.~\ref{fig:greif-soft-xte}, but for the
BPS+$\chi$EFT(stiff)+CS construction.  Panels (a) and (b) use representative hadronic branches of
moderate-$M_{\rm TOV}$ ($2.701\,\Msun$) and highest-$M_{\rm TOV}$ ($2.851\,\Msun$), with corresponding EOS parameters
$(a_1,a_2,a_3,a_4,a_5,a_6,a_7)=(2.0,10,4,3,1.4,-0.188,0.85)$ and
$(1.5,6,12,3,1.4,-1.475,0.7)$, respectively.
Triangles and color-bar marks follow Fig.~\ref{fig:hebeler-xte}.}
  \label{fig:greif-stiff-xte}
\end{figure*}

\twocolumngrid

\subsection{Comparison between the PP and CS representations}
\label{subsec:pp-cs-comparison}
As a robustness check of the main picture, the CS-parametrization scans in Figs.~\ref{fig:greif-soft-xte} and~\ref{fig:greif-stiff-xte} repeat
the same slow stable construction with a different high-density core representation.  We see a general agreement. A mild difference appears in the stiff $\chi$EFT matching case, where the representative branch in Fig.~\ref{fig:greif-stiff-xte}(a) does not realize scenario 1, reflecting the somewhat stronger sensitivity of this parametrization to the crust-core matching.  The overall agreement indicates that the slow stable mechanism is not an artifact of the piecewise-polytropic extension.
The CS sequences carry the same $n_c=n_{\rm pQCD}$ triangles, with
$\bigl(\Delta\eps/\eps_{\rm trans}\bigr)_{\rm crit}\simeq21$--$24$
(Sec.~\ref{subsec:pqcd}).

\section{Summary and Outlook}
\label{sec:summary}

We have examined whether a conventional, crust-bearing neutron-star EOS with a strong first-order
hadron-quark transition can reproduce the compact-star tensions previously addressed with
self-bound hybrid-star models.  The main result is that this is possible, but only through a
two-stage structure.  The pure hadronic branch must first retain enough high-density support to
reach a GW190814-scale mass.  For EOSs that are relatively soft at intermediate density but stiff at
higher density, the same hadronic branch can also pass through the HESS J1731--347 region, while the
slow stable hybrid continuation generated after the Maxwell jump can intersect the XTE J1814--338
small-radius contour.  This is the all-at-once realization of scenario 1.  In scenario 2, the
hadronic branch no longer covers HESS J1731--347, but very large density jumps can still drive the
slow stable branch into the HESS region; in that case, XTE J1814--338 and HESS J1731--347 are reached
in different transition-strength regimes.

Earlier slow-conversion studies that reached the HESS or XTE regions
did not incorporate the GW190814-scale mass
requirement~\cite{Mariani2024HESS,LaskosPatkos2025XTE,Mariani:2025nud}.
Both of our scenarios do: the hadronic branch must first support a
GW190814-scale mass, after which scenario~1 also covers HESS on that
hadronic branch and XTE on the slow continuation with one EOS, while
scenario~2 assigns XTE and HESS to different jump-strength regimes.
Other small-radius proposals---disconnected twins, Bayesian strong
transitions, self-bound stars, or an extra dark
component~\cite{LaskosPatkos2024HESS,Zhou:2026exw,Dong:2026kpb,Yang:2024ycl,Zhang2026SelfBound}---
need not rely on a postturning-point slow branch of a conventional
crust-bearing star; here the conversion timescale is part of the resolution.

The mechanism is therefore not characterized by a large maximum mass alone.  Its diagnostic feature
is the coexistence, within one EOS construction, of a high-mass hadronic branch and a compact
slow stable hybrid continuation.  In the polytropic benchmark with $\csssq=1$, the XTE
J1814--338 intersections occur for transitions near $M_t\simeq2.5$--$2.6\,\Msun$ and density jumps
of order a few times $\eps_{\rm trans}$, with similar windows in the CS-parametrization comparison.
Smaller jumps leave the hybrid sequence too close to the hadronic branch, whereas much larger jumps
bend it past the XTE region and instead become relevant for HESS-like compact configurations.  The
branch length is controlled jointly by the transition mass, transition strength, and the posttransition stiffness:
increasing $M_t$, decreasing $\Delta \eps/\eps_{\rm trans}$ or lowering $\csssq$ shortens the slow stable continuation and makes the
low-mass, small-radius XTE region harder to reach.

In the systematic classification of Mariani et al.~\cite{Mariani:2026xdh},
the viable continuations found here are late-transition type-W (waterfall)
branches.  Their result that larger density jumps and stiffer hadronic and
quark sectors favor extended type-W branches is consistent with our finding
that moderate jumps reach XTE, much larger jumps extend toward HESS, and
lowering $c_{s,\rm QM}^2$ shortens the continuation.  The analyses are
complementary: Mariani et al. mapped the morphology and normalized branch
length broadly, whereas our $M_{\rm TOV}>2.55\,\Msun$ selection and
$M_t=2.51$--$2.60\,\Msun$ transitions test this morphology in the more
extreme high-mass regime motivated by GW190814 and against the specific
HESS/XTE constraints.

The observational uncertainties associated with HESS J1731--347,
XTE J1814--338, and GW190814 affect the two scenarios differently.  A shifted
HESS contour would remove scenario~1's all-at-once advantage while leaving
the high-mass hadronic plus XTE continuation intact, and would mainly remove
scenario~2's motivation for the very-large-jump family.  An upward revision
of the XTE radius would weaken the need for an ultracompact slow continuation
in both scenarios.  If the GW190814 secondary is a black
hole~\cite{Tews2021GW190814,LVK2026GWTC5Population},
the $M_{\rm TOV}>2.55\,\Msun$ cut is no longer observationally
required, and the near-causal hadronic branches it selected become a stress
test rather than a data-driven necessity.

The obtained branch-resolved tidal-deformability calculations support the same picture.  The moderate-$M_{\rm TOV}$ hadronic branches remain below the usual
GW170817 upper-bound scale, while the highest-$M_{\rm TOV}$ hadronic examples are close to or
slightly above it.  A useful next step is therefore to revisit the refined tidal-deformability
upper and lower bounds in the literature with strong first-order phase transitions and slow stable
branches included explicitly, rather than importing those bounds as fixed cuts from analyses based
on pure hadronic-matter sequences or quasiuniversal relations.

A more definite determination on the existence of the slow stable branch would require microscopic calculations of nucleation and conversion timescales, which can connect the
slow-conversion assumption to phase-boundary dynamics. Besides, a Bayesian analysis over the hadronic and quark-matter EOS space can systematically
quantify how strongly the transition window identified here is supported by current and future
data. A complementary extension is to test these branch-resolved EOSs directly against GW spectra and waveforms, which can carry observable imprints of nonhadronic degrees of freedom and additional dense-matter components beyond static $M$--$R$ and $\Lambda(M)$ diagnostics~\cite{Alford:2014jha,Han:2019dpq,Miao:2023jqe,Hong:2024srz,Hong:2026ugt,Pereira:2025xsi,Counsell:2025hcv,Gao:2026wje}. We leave these open problems for future studies.

\section*{ACKNOWLEDGMENTS}

C.Z. is supported by the Fundamental Research Funds for the
Central Universities and the Shanghai Pujiang Program. We are grateful for the stimulating discussions during the 2nd Workshop on Dense Matter Equation of State and Frontiers in Neutron Star Physics (DMNS 2026) held at the T.D. Lee Institute, Shanghai Jiao Tong University, during which part of the work was conducted.

\section*{DATA AVAILABILITY}

There are no publicly available research data or software supporting this manuscript. Requests for further information or data should be sent to the authors.

\appendix

\setcounter{section}{0}
\renewcommand{\thesection}{\Alph{section}}
\setcounter{secnumdepth}{1}

\section{Stellar structure and radial stability equations}
\label{app:stability-equations}

For completeness, we record the equations used for the stellar backgrounds and the radial-mode
stability test.  The static spherically symmetric metric is
\begin{equation}
ds^{2}=-e^{2\Phi(r)}dt^{2}+e^{2\Psi(r)}dr^{2}
+r^{2}(d\theta^{2}+\sin^{2}\theta\,d\phi^{2}),
\end{equation}
with $e^{-2\Psi(r)}=1-2m(r)/r$.  For a given EOS, the stellar structure follows from the
Tolman-Oppenheimer-Volkov (TOV) equations~\cite{Tolman1939,Oppenheimer1939}
\be
\begin{aligned}
\frac{dP}{dr}
&=-\frac{[m(r)+4\pi r^{3}P(r)][\eps(r)+P(r)]}{r[r-2m(r)]}, \\
\frac{dm}{dr}
&=4\pi \eps(r)r^{2}, \\
\frac{d\Phi}{dr}
&=-\frac{1}{\eps(r)+P(r)}\frac{dP}{dr},
\end{aligned}
\label{eq:tov}
\ee
with $P(R)=0$ and $m(R)=M$ defining the stellar radius and gravitational mass.

Radial perturbations are written in terms of the displacement
$\xi=\Delta r/r$ and pressure perturbation $\Delta P$, with harmonic time dependence
$e^{i\omega t}$.  The first-order system is~\cite{Chandrasekhar1964,Chanmugam1977,Vath1992,
Gondek1997,Vasquez2010}
\begin{eqnarray}
\frac{d\xi}{dr}&=&\V(r)\xi+\W(r)\Delta P, \label{eqXi} \\
\frac{d\Delta P}{dr}&=&X(r)\xi+Y(r)\Delta P, \label{eqDP}
\end{eqnarray}
where
\begin{eqnarray}
\V(r)&=&-\frac{3}{r}-\frac{1}{P+\eps}\frac{dP}{dr},\\
\W(r)&=&-\frac{1}{r\Gamma P},\\
X(r)&=&\omega^{2}e^{2\Psi-2\Phi}(P+\eps)r \nonumber\\
&&-4\frac{dP}{dr}
+\left(\frac{dP}{dr}\right)^{2}\frac{r}{P+\eps} \nonumber\\
&&-8\pi e^{2\Psi}(P+\eps)Pr,\\
Y(r)&=&\frac{1}{P+\eps}\frac{dP}{dr}
-4\pi(P+\eps)r e^{2\Psi},
\end{eqnarray}
and $\Gamma=(\eps+P)P^{-1}dP/d\eps$ is the adiabatic index.  We use
$(\Delta P)_{r=0}=-3(\xi\Gamma P)_{r=0}$, $\xi(0)=1$, and the surface condition
$(\Delta P)_{r=R}=0$.  The stability boundary is identified by the point at
which the squared fundamental-mode frequency, $\omega_0^2$, vanishes.  For rapid phase conversion, the interface conditions are
$[\Delta P]^+_-=0$ and
$[\xi-\Delta P/(rP')]^+_-=0$.  For slow conversion, used in the main analysis, the  interface conditions are
$[\xi]^+_-=0$ and $[\Delta P]^+_-=0$; these junction conditions and their dynamical interpretation follow
Refs.~\cite{Haensel1989,Pereira2018}.

For comparisons with gravitational-wave observations, the dimensionless tidal deformability $\Lambda=2k_2/(3C^5)$ can be computed, where $C=M/R$ is the compactness and $k_2$ is the Love number characterizing the stellar response to external perturbations~\cite{Hinderer2008,Hinderer2010,Postnikov2010}.
The Love number $k_2$ is determined by solving a differential equation for $y(r)$ \cite{Postnikov2010} concurrently with the TOV equations in Eq.~\eqref{eq:tov}, using the boundary condition $y(0)=2$. For hybrid stars, the matching condition $y(r_{d}^+) - y(r_{d}^-) = -4\pi r_{d}^3 \Delta \eps_d /(m(r_{d})+4\pi r_{d}^3 P(r_{d}))$ must be applied at the radius $r_d$ where $P(r_d)=P_{\rm trans}$ and an energy density jump $\Delta \eps_d$ occurs~\cite{damour2009relativistic,takatsy2020comment}.

\section{Accelerated end points search: the residual method}
\label{app:method1}

The slow stable end points used in the main analysis are the marginal-stability points at which the
fundamental radial mode frequency vanishes, $\omega_0^2=0$.  A standard implementation treats
$\omega^2$ as the eigenvalue at fixed central pressure $P_c$: for each trial star one integrates
Eqs.~\eqref{eqXi} and~\eqref{eqDP} while shooting in $\omega^2$ until the surface condition
$\Delta P(R)=0$ is satisfied, thereby obtaining $\omega_0^2(P_c)$, and then scans $P_c$ until
$\omega_0^2(P_c^*)=0$.  This nested search---an inner eigenfrequency solve at every $P_c$, wrapped
in an outer sequence or root search in $P_c$---is unnecessarily expensive when the only target is
the stability end point itself.

We introduce the \emph{residual method} (RM), a rapid end point search method that removes the inner loop
by exploiting the fact that the marginal mode is a static mode.  Rather than reconstructing
$\omega_0^2(P_c)$ at every trial central pressure, the RM sets $\omega^2=0$ from the outset in
Eqs.~\eqref{eqXi} and~\eqref{eqDP} and locates the end point from the root of the surface
residual
\begin{equation}
  F(P_c)\equiv \Delta P(R;P_c,\omega^2=0),
  \label{eq:method1-residual}
\end{equation}
the Lagrangian pressure perturbation evaluated at the stellar surface after a single
zero-frequency integration.  The simplification is immediate in the coefficient $X(r)$:
the $\omega^2 e^{2\Psi-2\Phi}(P+\eps)r$ term vanishes, and the perturbation system becomes a
purely static first-order system driven only by the background profile.  For each trial $P_c$, the RM
therefore performs a single pass:
\begin{enumerate}
  \item integrate the TOV equations outward to the surface $P(R)=0$;
  \item integrate Eqs.~\eqref{eqXi} and~\eqref{eqDP} with $\omega^2=0$, using the regular central
  conditions $\xi(0)=1$ and $(\Delta P)_{r=0}=-3(\xi\Gamma P)_{r=0}$;
  \item at a first-order phase transition, apply the slow-conversion junction conditions
  $[\xi]^+_-=0$ and $[\Delta P]^+_-=0$ by keeping both perturbation variables continuous while
  switching the EOS branch in the background coefficients;
  \item evaluate $F(P_c)$ at the surface.
\end{enumerate}
The end point central pressure $P_c^*$ is the root of the one-dimensional equation
\begin{equation}
  F(P_c^*)=0,
  \label{eq:method1-root}
\end{equation}
which may be located with any robust scalar root finder once a bracketing interval has been
identified.  Each evaluation of $F(P_c)$ requires only one background integration and one
zero-frequency perturbation integration, so the marginal point is obtained directly rather than
by reconstructing the full $\omega_0^2(P_c)$ curve.

Conceptually, the RM inverts the usual eigenvalue problem,
\begin{equation}
  \text{standard:}\quad \omega^2=\omega^2(P_c),
  \qquad
  \text{RM:}\quad P_c=P_c(\omega^2=0),
  \label{eq:method1-inversion}
\end{equation}
replacing nested eigenfrequency shooting with a one-dimensional root search in $P_c$ on the
zero-frequency slice of Eqs.~\eqref{eqXi} and~\eqref{eqDP}.

For slow phase conversions, the RM is the efficient route to the end point.  For rapid conversion,
 in contrast, the stability limit of an equilibrium sequence usually coincides with the mass turning
point, $\partial M/\partial P_c=0$, and a full radial-mode solve is not needed if that dynamical
limit is assumed.  The present paper focuses on the slow limit, where $\partial M/\partial P_c<0$
does not by itself imply instability and the RM therefore supplies the end point location used in the
scans.

\section{Hadronic sound-speed profiles}
\label{app:sound-speed-profiles}

Figure~\ref{fig:app-hebeler-cs2-vs-pc} shows the central sound speed along the
pure-hadronic PP branches used in the slow stable hybrid scans.  The curves are evaluated from the analytic polytropic
thermodynamic relation, rather than by interpolating the sparse stellar
sequence, so the vertical line segments mark genuine sound-speed jumps at
piecewise-polytropic interfaces.  Once the raw PP core reaches the causal
limit, the causal extension of Ref.~\cite{Hebeler2013} produces the expected $c_s^2=1$ plateau.

Very narrow rectanglelike features occur
when two such jumps, or a jump and the causal-extension onset, are separated by
only a small finite pressure interval; they reflect closely spaced continuous
PP segments rather than any pressure discontinuity.

Hadronic stars with $M>2.51 \, M_\odot$ need not all lie on the $c_s^2=1$ plateau: in a small number of
softer $\chi$EFT-matching background branches the $2.51\,\Msun$ configuration is reached at a central
pressure slightly below the causal-extension onset.  These off-plateau stars
nearly overlap because the corresponding branches share the same lower-density
PP parameters and differ only in $\rho_{23}$, which is not yet probed at the
$2.51\,\Msun$ central pressure.

The same $\Gamma_2=0$ cases excluded from the mass-radius plots are also
excluded here, since those entries mimic a first-order transition inside the
hadronic PP core rather than the separate CSS transition used in this work.

Despite the near-causal central sound speeds at the high-mass end,
the central densities on these pure-hadronic sequences remain moderate up to
the transition masses marked in the figures: for every retained PP and CS
branch that reaches $M_t=2.51$ or $2.60\,\Msun$, one finds
$n_c\lesssim5n_0$ there, so $n_c<n_{\rm pQCD}=40n_0$ throughout
$M\le M_t$.  Crossing into the $n_{\rm pQCD}$ regime therefore requires the
posttransition CSS core discussed in Sec.~\ref{subsec:pqcd}, not the
pretransition hadronic profiles shown here.

Figure \ref{fig:app-greif-cs2-vs-pc} gives the same central-sound-speed
diagnostic for the CS-parametrization comparison used in Figs.~\ref{fig:greif-soft-xte}
and~\ref{fig:greif-stiff-xte}.  In this case, the core sound speed is evaluated
directly from Eq.~\eqref{eq:cs2}, with the integrated $P(\eps)$ relation used
only to map the stellar central pressure to the corresponding central energy
density.  The resulting curves are smooth apart from the causal clipping to
$c_s^2=1$, in contrast to the jump structure of the PP benchmarks.

The near-causal hadronic sound speeds shown in Figs. \ref{fig:app-hebeler-cs2-vs-pc} and \ref{fig:app-greif-cs2-vs-pc}  should be compared with
conformality-motivated EOS inferences with their underlying assumptions in
mind.  Recent conformality-motivated studies connect low-density nuclear input
and astrophysical constraints to high-density QCD information within ordinary
stable-sequence EOS ensembles~\cite{Annala:2019puf,Annala:2023cwx,Gorda2023QCDInput,Fujimoto:2022ohj}.
Their high-density softening is therefore not purely fixed by current
neutron-star data, but is also influenced by the pQCD end point or likelihood,
which tends to reduce the pressure and sound speed at the highest densities
\cite{Gorda2023QCDInput}; the associated sound-speed peak can then be
interpreted as compatible with an approach of the trace anomaly toward
conformality~\cite{Fujimoto:2022ohj}.  This does not mean, however, that the
conformal value is a robust upper bound at neutron-star densities.  Tews et
al.~\cite{Tews2018} emphasize that pQCD constrains the asymptotically
high-density regime, far above densities reached in neutron-star cores, and
therefore allow $c_s^2$ to exceed the conformal value while retaining only the
causal bound.  Low-density nuclear input together with two-solar-mass pulsars
then points to tension with imposing $c_s^2\le 1/3$ throughout the star
\cite{Bedaque2015,Tews2018}.  Once this conformal bound is relaxed, the sound
speed typically rises rapidly around $\sim2n_0$, can approach the causal limit
in many viable parametrizations, and has a lower bound on $c_{s,\max}^2$ that
increases with the maximum mass to be supported~\cite{Tews2018}.

\par\onecolumngrid
\begingroup
\setlength{\intextsep}{10pt}
\begin{figure}[!h]
  \hsize=\textwidth
  \linewidth=\textwidth
  \centering
  \begin{minipage}[t]{0.49\textwidth}
    \centering
    \includegraphics[width=\linewidth]{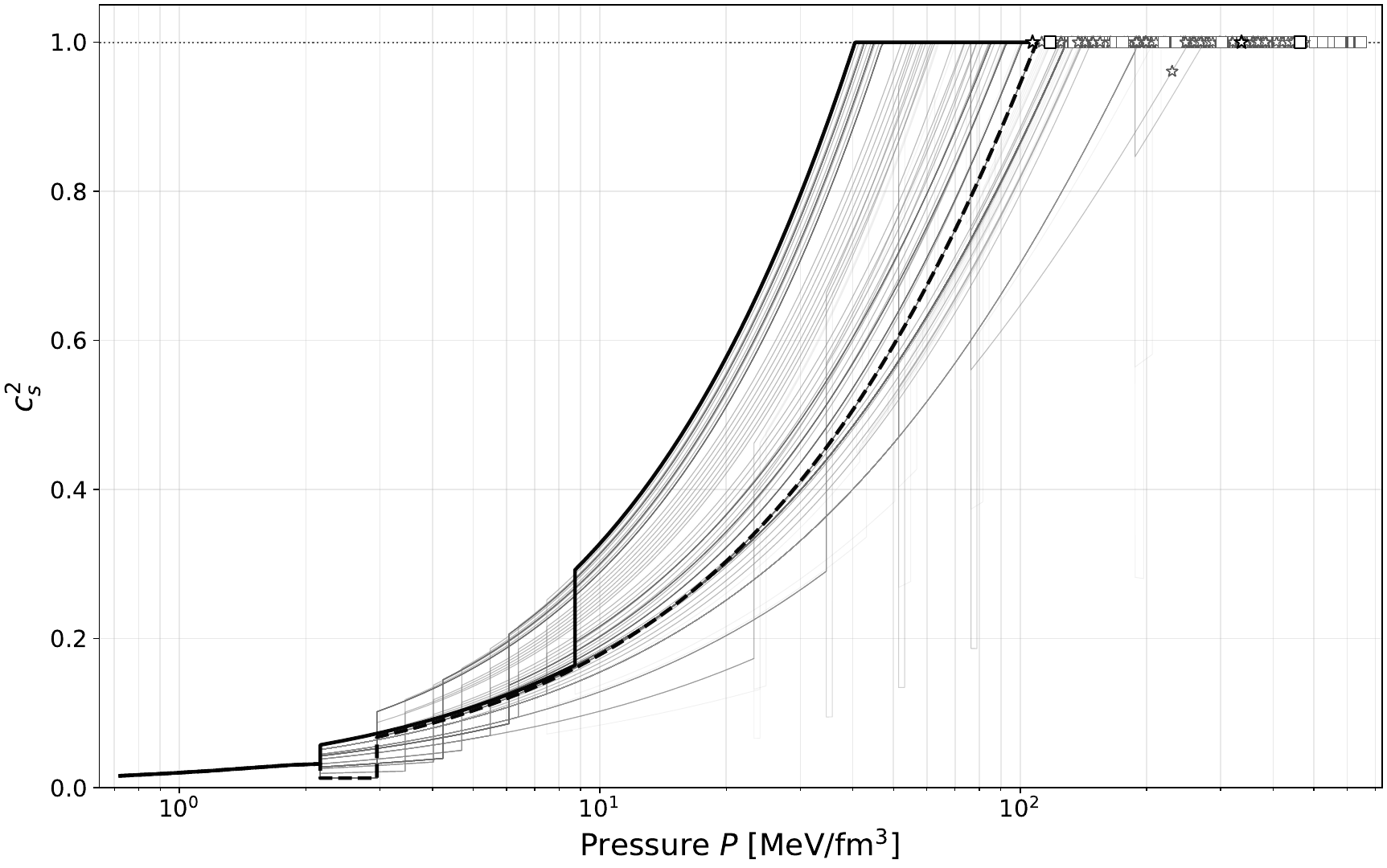}\\[-0.4em]
    \textbf{(a)}
  \end{minipage}\hfill
  \begin{minipage}[t]{0.49\textwidth}
    \centering
    \includegraphics[width=\linewidth]{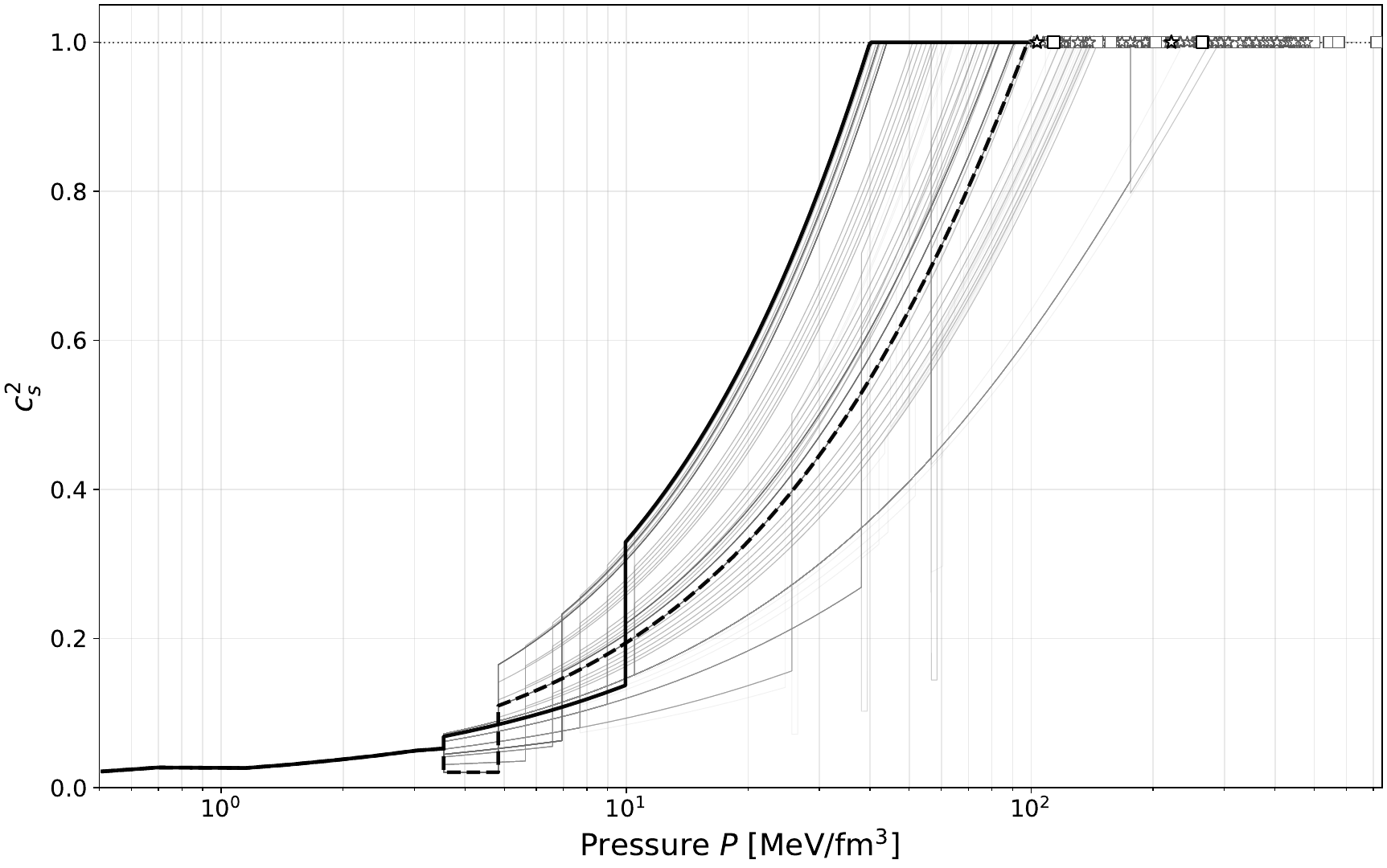}\\[-0.4em]
    \textbf{(b)}
  \end{minipage}
  \caption{Central sound speed $c_s^2(P_c)$ for the PP benchmark pools of
  Figs.~\ref{fig:hebeler-xte} and~\ref{fig:hebeler-xte-stiff}, using the same
  style conventions.  Thick black dashed and
  solid curves mark the representative moderate- and highest-$M_{\rm TOV}$
  hadronic branches, respectively; open stars and squares denote
  $M_t=2.51\,\Msun$ and $2.60\,\Msun$.
  At these $M_t$ markers, $n_c\lesssim5n_0$ on all retained branches
  (Sec.~\ref{subsec:pqcd}).}
  \label{fig:app-hebeler-cs2-vs-pc}
\end{figure}
\FloatBarrier
\endgroup
\begin{figure*}[t]
  \hsize=\textwidth
  \linewidth=\textwidth
  \centering
  \begin{minipage}[t]{0.49\textwidth}
    \centering
    \includegraphics[width=\linewidth]{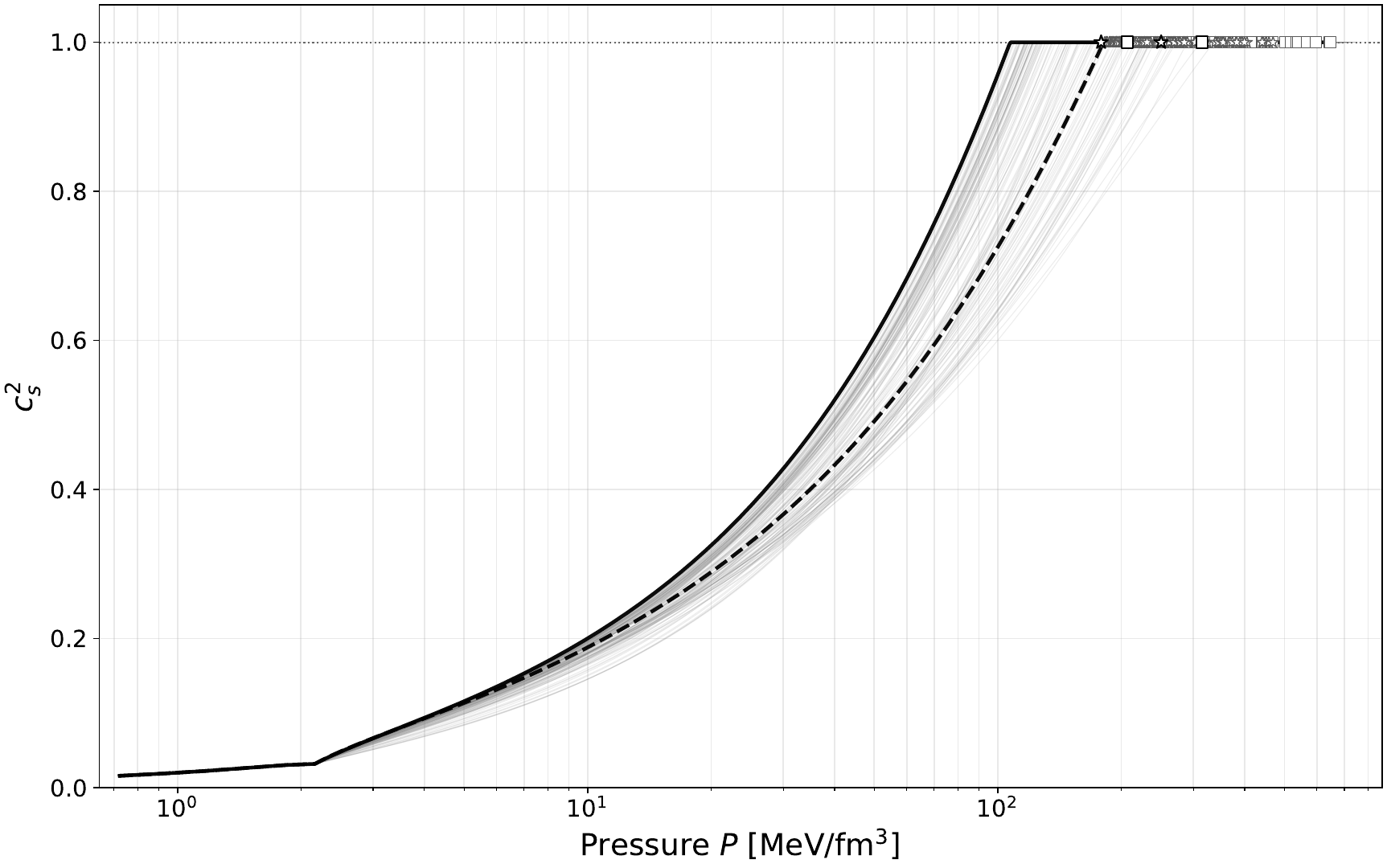}\\[-0.4em]
    \textbf{(a)}
  \end{minipage}\hfill
  \begin{minipage}[t]{0.49\textwidth}
    \centering
    \includegraphics[width=\linewidth]{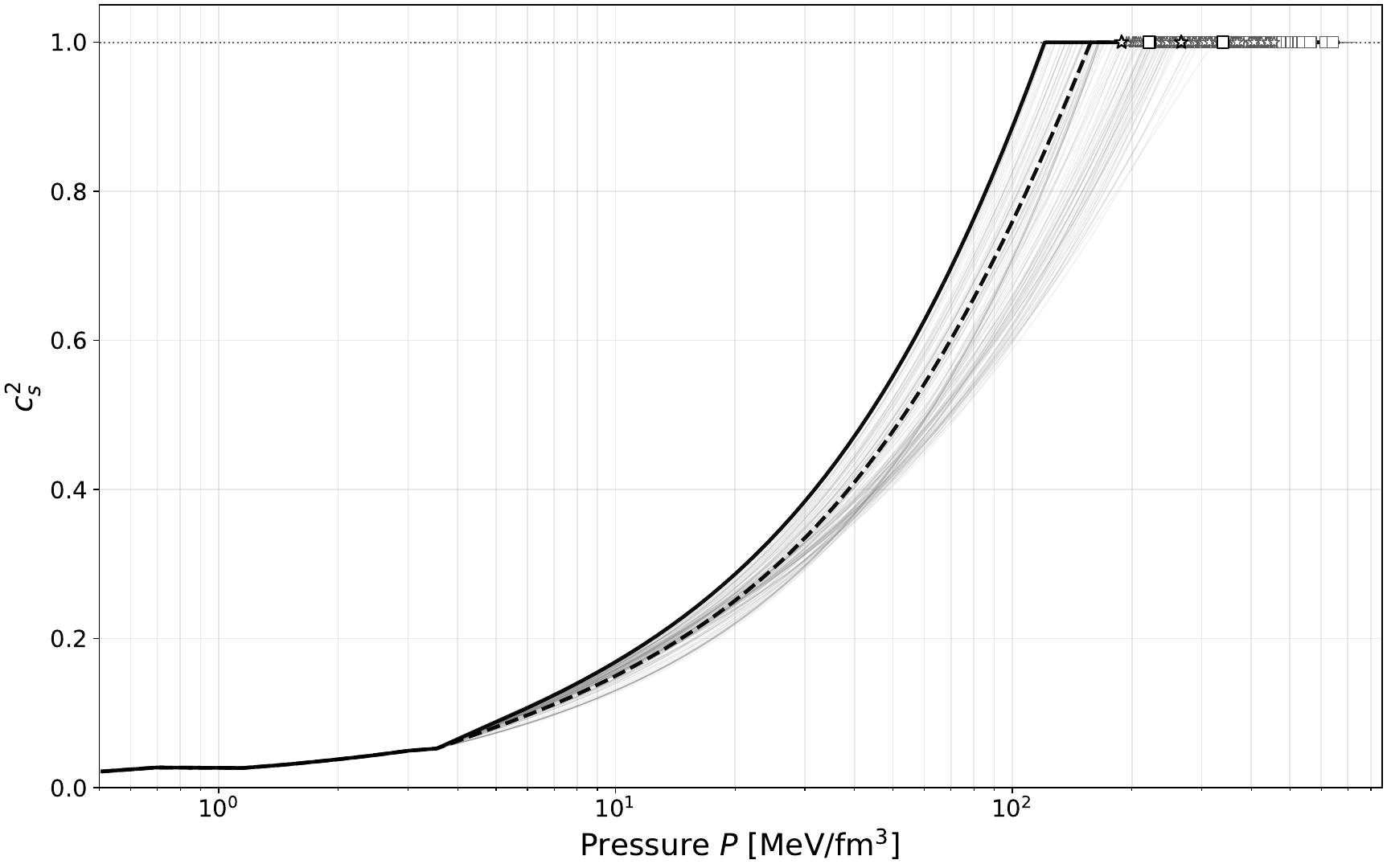}\\[-0.4em]
    \textbf{(b)}
  \end{minipage}
  \caption{Central sound speed $c_s^2(P_c)$ for the CS comparison of
  Figs.~\ref{fig:greif-soft-xte} and~\ref{fig:greif-stiff-xte}, using the same
  style conventions.  Thick black dashed and
  solid curves mark the representative moderate- and highest-$M_{\rm TOV}$
  hadronic branches, respectively; open stars and squares denote
  $M_t=2.51\,\Msun$ and $2.60\,\Msun$.
  As in Fig.~\ref{fig:app-hebeler-cs2-vs-pc}, $n_c\lesssim5n_0$ at
  these $M_t$ markers on all retained branches.}
  \label{fig:app-greif-cs2-vs-pc}
\end{figure*}

\twocolumngrid
\noindent

\bibliography{references_v5}

\end{document}